\documentclass{aa}  

\usepackage{natbib}
\usepackage{graphicx}
\usepackage{txfonts}
\usepackage[pdftex]{hyperref}
\hypersetup{urlcolor=black, citecolor=blue, linkcolor=blue, colorlinks=true}  % Colours hyperlinks in blue, but this can be distracting if there are many links.
\usepackage{amsmath}
\usepackage{amssymb}
\usepackage{bm}
\usepackage{balance}

\begin{document} 
    \title{XMMPZCAT: A catalogue of photometric redshifts for X-ray sources}
    \author{A. Ruiz
            \inst{1}
        \and
            A. Corral
            \inst{2}
        \and
            G. Mountrichas
            \inst{1}            
        \and
            I. Georgantopoulos
            \inst{1}
    }

    \institute{  
        Institute for Astronomy, Astrophysics, 
        Space Applications, and Remote Sensing (IAASARS), 
	    National Observatory of Athens,
	    15236 Penteli, Greece \\
        \email{ruizca@noa.gr}	      
      \and   
        Instituto de Física de Cantabria (UC-CSIC),
        Av. de los Castros s/n, 39005, Santander, Spain
    }

   \date{Received March 28, 2018; accepted July 10, 2018}

% \abstract{}{}{}{}{} 
% 5 {} token are mandatory
 
  \abstract
   {}
   {The third version of the XMM-\textit{Newton} serendipitous catalogue (3XMM), 
    containing almost half million sources, is now the largest X-ray catalogue. 
    However, its full scientific potential remains untapped due to 
    the lack of distance information (i.e. redshifts) for the majority of 
    its sources. Here we present XMMPZCAT, a catalogue of photometric redshifts
    (photo-z) for 3XMM sources.}
   {We searched for optical counterparts of 3XMM-DR6 sources outside the Galactic 
    plane in the SDSS and Pan-STARRS surveys, with the addition of near- (NIR) and 
    mid-infrared (MIR) data whenever possible (2MASS, UKIDSS, VISTA-VHS, and AllWISE). 
    We used this photometry data set in combination with a training sample of 5157 
    X-ray selected sources and the MLZ-TPZ package, a supervised machine learning
    algorithm based on decision trees and random forests for the calculation of
    photo-z.}
   {We have estimated photo-z for 100\,178 X-ray sources, about 50\% 
   of the total number of 3XMM sources (205\,380) in the XMM-\textit{Newton} 
   fields selected to build this catalogue (4208 out of 9159). The accuracy 
   of our results highly depends on the available photometric data, with a 
   rate of outliers ranging from 4\% for sources with data in the optical+NIR+MIR, 
   up to $\sim 40\%$ for sources with only optical data. We also addressed 
   the reliability level of our results by studying the shape of the photo-z 
   probability density distributions.}
   {}


   \maketitle
%
%________________________________________________________________

\section{Introduction}
\label{sec:intro}
The third version of the XMM-\textit{Newton} serendipitous catalogue, data 
release six, (3XMM-DR6) contains about 500\,000 unique sources covering a 
total area of 1000\,deg$^2$ on the sky. Two thirds of these are located at 
high Galactic latitude, |b|>20 deg. Recently, the catalogue has been enriched 
with added-value products thanks to the XMMFITCAT \citep{xmmfitcat}, ARCHES 
\citep{arches} and EXTraS \citep[e.g.][]{Pizzocaro16} projects. The first 
provides the spectral fit inventory of 157\,000 sources with the highest 
photon statistics while the second is providing multi-wavelength positional 
cross-matches using Bayesian statistics. The third project will provide the 
characterisation of X-ray variable sources. 
 
Although the vast majority of these nearly 500\,000 sources are expected to 
be active galactic nuclei (AGN), the huge potential of the 3XMM catalogue 
remains practically untapped because most of the sources lack redshift 
information. Spectroscopy is expensive and challenging for such a large 
sample. Therefore, photometric redshifts (photo-z) is the only feasible 
way to estimate distances for half a million sources. However, photo-z 
are subject to systematics and higher uncertainties compared to spectroscopic 
redshifts (spec-z). 

Many photo-z estimation methods have been developed in pursuit of deriving the 
most accurate photo-z. These methods can be divided into two main categories: 
template-fitting \citep[e.g.][]{Bolzonella00} and machine-learning 
\citep[e.g.][]{DIsanto18} techniques. 

Among the variety of methods developed using template-fitting techniques we 
can mention for instance Bayesian photometric redshifts \citep[BPZ;][]{Benitez00} 
or Easy and Accurate photo-Z from Yale \citep[EAZY;][]{Brammer08}. In the case of 
machine learning techniques, a growing number of methods have been published in
recent years, using different machine learning algorithms, from neural networks,
for example artificial neural network \citep[ANNz;][]{Collister04,Lahav12}, to 
random forest techniques, for example, Trees for Photo-Z \citep[TPZ;][]{tpz}.
Each of these methods has its own advantages and disadvantages (see \citealt{Abdalla11,Beck17} 
for detailed comparisons among the various photo-z estimation methods). Mixed 
techniques are also used, where spectral energy distribution (SED) fitting methods 
rely on a previous machine-learning classification to select the most adequate 
set of templates \citep[e.g.][]{Fotopoulou16}.

Supervised machine-learning techniques, also known as empirical methods, 
require a spectroscopic sample to train an algorithm. Then the algorithm 
is applied to a dataset with available photometry to estimate photometric 
redshifts. These methods have been extensively applied on galaxy samples 
\citep[e.g. SDSS,][]{Beck16} and optical QSOs \citep[e.g.][]{Ball07,Ball08,DIsanto18}. 
For X-ray AGN, though, only SED fitting techniques had been used 
\citep{Salvato09,Hsu14}, up until recently. This was due to lack of large 
enough X-ray spectroscopic training sets. \citet{Mountrichas17} presented a 
catalogue with $\sim 1000$ X-ray sources in the X-ATLAS field using, for the 
first time, a machine learning technique \citep[TPZ;][]{tpz}. Their training, 
spectroscopic sample consists of 5157 sources primarily in the XMM-XXL field \citep{Liu16,Menzel16}.

Here we present the XMM-\textit{Newton} Photo-Z CATalogue (XMMPZCAT). Using 
the training set presented in \citet{Mountrichas17}, we estimate photometric 
redshifts for $\sim 100\,000$ 3XMM sources with at least optical photometry 
available (SDSS or Pan-STARRS). We provide estimations of the accuracy of 
our photo-z estimations as well as the percentage of outliers.

The structure of the paper is as follows: in Sect.~\ref{sec:data} we describe 
the 3XMM sample as well as the catalogues used to obtain photometric information 
for our X-ray sources. In Sect.~\ref{sec:crossmatch} we present the methods 
used for the cross-match among the various datasets and in Sect.~\ref{sec:photoz} 
we briefly describe the TPZ algorithm and the training sample. The results are 
presented in Sect.~\ref{sec:results}, while we summarise the main conclusions 
of our analysis in Sect.~\ref{sec:conclusions}.

\section{Data}
\label{sec:data}
Optical photometry is needed to derive photometric redshifts, so
we explored several wide-angle surveys to maximise the number of
counterparts of our X-ray sources. We also complemented these data
with photometry in the near- and mid-infrared to increase the
accuracy of the photometric redshifts. A brief description of all
catalogues we used is presented below.

\subsection{3XMM}
\label{subsec:3xmm}
3XMM-DR6 catalogue \citep{3xmm} was released in July 2016. It contains 
9160 observations, covering an energy interval from 0.2~keV to 12~keV. 
The net sky area observed is $\sim$ 1032\,deg$^2$. 468\,440 unique X-ray 
sources are included in the catalogue with a median flux in the total 
energy band of  $\sim 2.4 \times 10^{-14}~\mathrm{erg\,cm^{-2}\,s^{-1}}$.

\subsection{SDSS}
\label{subsec:sdss}
Data release 13 of the Sloan digital sky survey \citep[SDSS-DR13;][]{sloan13} 
is the first data release of the fourth phase of the SDSS and covers 
14\,555\,deg$^2$, more than one-third of the entire celestial sphere. This 
release does not include new or updated photometric information beyond that
already included in the SDSS-DR9 \citep{sloan9}, but the imaging data have 
been photometrically re-resolved and recalibrated. The catalogue contains 
about 500 million unique, primary sources providing optical photometry for 
about 95\% of the point sources in five bands: u, g, r, i, z. The magnitude 
limit is $r_{AB}=22.2$\,mag. Among the various measures of magnitude that
offers the SDSS, we used the composite model magnitudes (cModelMag). These
magnitudes are suitable for both extended and point-like sources.

\subsection{Pan-STARRS}
\label{subsec:pstarrs}
Pan-STARRS1 \citep[PS1;][]{pstarrs1db} is the first data release from the 
panoramic survey telescope and rapid response system \citep[Pan-STARRS;][]{pstarrs1survey}. 
Pan-STARRS uses a 1.8 metre telescope to image the sky. It exploits the 
combination of relatively small mirrors with very large digital cameras to 
observe the entire available sky several times each month. 3$\pi$ of the 
sky are covered north of declination -30 deg, in five broad-band filters: 
g, r, i, z, y. The magnitude limits reach g=23.3, r=23.2, i=23.1, z=22.3, 
and y=21.3.

Whenever available, we used the stack photometry (i.e. the magnitudes 
estimated using the stacking of all available PS1 observations of the source), 
otherwise we used the mean photometry. For objects classified as point-like 
by the PS1 pipeline, we used the corresponding PSF magnitudes. For extended
objects, we used the Kron magnitudes.

\subsection{AllWISE}
\label{subsec:wise}
NASA's wide-field infrared survey explorer \citep[WISE;][]{wise} mapped the 
sky in the mid-infrared (MIR) at 3.4, 4.6, 12, and 22 $\mu$m (W1, W2, W3, W4). 
WISE achieved 5$\sigma$ point source sensitivities better than 0.08, 0.11, 1 
and 6 mJy in unconfused regions on the ecliptic in the four bands. Sensitivity 
improves towards the ecliptic poles due to denser coverage and lower zodiacal 
background. The AllWISE source catalogue \citep{allwise} contains the attributes 
for over 747 million objects detected at SNR>5 in at least one band in the 
combined exposures of the atlas intensity images. 

\subsection{2MASS}
\label{subsec:2mass}
The 2 micron all sky survey \citep[2MASS;][]{Skrutskie06} uniformly scanned 
the entire sky in three near-infrared (NIR) bands, J, H, K$_S$. The magnitude 
limits are J=15.8/15.0, H=15.1/14.3 and K$_S=14.3/13.5$ for point-like or extended 
sources, correspondingly. We used the 2MASS point-source catalogue \citep{2MASS}, 
which includes over 300 million objects.

\subsection{UKIDSS}
\label{subsec:ukidss}
The UKIRT \citep{Casali07} infrared deep sky survey \citep[UKIDSS;][]{ukidss}, 
successor to 2MASS, started in 2005 and has surveyed 7500\,deg$^2$ on the Northern 
sky in YJHK down to a magnitude limit of K=18.4 (Vega) in its shallowest parts 
(UKIDSS-LAS, 4000\,deg$^2$). This depth is three magnitudes deeper than 2MASS. 
We have used the data release ten of UKIDSS-LAS, the most recent release (November 
2014) publicly available. Only J, H and K magnitudes were used in our cross-match.

\subsection{VISTA-VHS}
\label{subsec:vista}
The visible and infrared survey telescope for astronomy \citep[VISTA;][]{ukidss} 
is an ESO's 4 metre class wide field survey telescope for the southern hemisphere, 
equipped with a near infrared camera. The VISTA-VHS \citep[VISTA hemisphere survey;][]{McMahon13} 
is one of the six large public surveys conducted by VISTA to image the entire 
southern hemisphere of the sky ($\sim 20\,000~\mathrm{deg}^2$), with the 
exception of the areas already covered by other vista surveys surveys, in J 
and K$_S$ bands, plus Y and H bands for the high Galactic latitude ($|b| > 
30$~deg) sky. VISTA-VHS has a depth advantage of over four magnitudes relative 
to previous NIR surveys like 2MASS or DENIS \citep{denis}, and $\sim 1.3$ 
magnitudes relative to the UKIDSS-LAS. The magnitude limits are J=20.2, H=19.4 
and K$_S=18.1$. We used data from VHS-DR4, the last data release publicly 
available (March 2017).

\section{Identification of counterparts}
\label{sec:crossmatch}
Our first step in building XMMPZCAT was the identification of optical 
counterparts for the X-ray sources in the 3XMM and, whenever possible, 
counterparts in the NIR and/or MIR bands. We looked for optical 
counterparts using two large area surveys: SDSS and PS1. For 
the SDSS, we used the multi-wavelength catalogues from the ARCHES 
project (see Sect.\ref{subsec:arches} below). For PS1, we did 
our own cross-matching of 3XMM and PS1 (see Sect.\ref{subsec:xmatch}), 
plus the NIR and MIR catalogues described above, using the cross-matching 
tool and techniques developed by ARCHES.

There are other tools that offer basic cross-matching algorithms
\citep[e.g. TOPCAT;][]{topcat}, or even more advanced techniques using 
`Bayes' factors' \citep{Budavari08} that combine astrometric and 
photometric information \citep{Georgakakis11}, but none of these
handle the statistics inherent to the crossmatching process in a 
fully coherent manner, particularly when combining the information 
from more than two catalogues. A proper statistical treatment is of
great importance when dealing with catalogues where the positional
uncertainties are significantly different, which is usually the case 
if X-ray catalogues are included. For example, the average error in 
the 3XMM is $\sim2$~arcsec, while in optical catalogues like SDSS
or Pan-STARRS is one or even two orders of magnitude lower.

The design of a crossmatching tool with such capabilities was one of 
the major goals of ARCHES. Recently, an alternative tool with similar
capabilities has been developed \citep[NWAY;][]{Salvato18}.

\subsection{3XMMe-SDSS (ARCHES)}
\label{subsec:arches}
The astronomical resource cross-matching for high energy 
studies (ARCHES) project is an European consortium created 
with the objective of building scientifically-validated 
spectral energy distributions for the many X-ray sources 
detected by the XMM-Newton observatory. To this end, ARCHES 
developed the enhanced 3XMM catalogue data (3XMMe) and 
new tools for the cross-correlation of extensive astronomical 
catalogues. See \citet{arches} for further details.

Among the main deliverables of the ARCHES project, two
multi-wavelength cross-matched catalogues were released. The 
base X-ray catalogue used in the cross-matching was 3XMMe. 
This catalogue is based on the 3XMM-DR5 catalogue and it was 
built by carefully removing detections that are considered 
to be of lower scientific reliability or quality, as well as 
fields that have been the subject of dedicated studies and/or
overlap with large, well known objects. As a result, 3XMMe 
is smaller than 3XMM-DR5, but composed of X-ray sources from 
XMM-\textit{Newton} observations with the highest quality.

The first of the multi-wavelength cross-matched catalogues is the
result of the cross-correlation of 3XMMe with GALEX-DR5 \citep{galexgr6}, 
UCAC4 \citep{ucac}, SDSS, 2MASS, AllWISE, the merge of SUMSS and 
NVSS \citep{Mingo16}, and the AKARI-FIS \citep{akarimaps} catalogues. 
The second one is similar, although smaller, since 2MASS is replaced 
by UKIDSS-LAS, which is not an all-sky survey. Probabilities of 
associations were computed, by using the \texttt{xmatch} tool (see 
Sect.~\ref{subsec:xmatch}), for all possible sets of candidates in 
the first five catalogues, whereas candidates from SUMSS, NVSS and 
AKARI-FIS were selected based on a $\chi^2$ criteria.

We were interested in finding counterparts in the optical (SDSS), NIR 
(2MASS or UKIDSS), and MIR (WISE). Therefore, we rejected all sources with
no counterparts in at least one of these catalogues. Moreover, we only 
kept counterparts with photometry in all the corresponding filters of each 
catalogue. For example, if an X-ray source has counterparts in the NIR 
and the optical, but the NIR source has no photometry in the J filter, we 
rejected the NIR counterpart and considered that the X-ray source has only 
an optical counterpart. If the optical source has no photometry in one or 
more SDSS filters, the X-ray source is not included in our selection.

We did a second filtering based on the probabilities of association given
in the ARCHES catalogue. However, since these probabilities  take into 
account all the catalogues where the counterparts were found, we had 
to derive the probabilities of associations for the cases we were interested
in, as a function of the ones provided in the ARCHES catalogues, that is, 
we estimated the marginal probabilities for the optical, optical+NIR, 
optical+MIR, and optical+NIR+MIR according to each case.

We then selected sources with probabilities of association larger than
68\%. The same X-ray source can appear multiple times in the ARCHES 
catalogue, associated with a different set of optical, NIR, and/or MIR 
counterparts. In those cases we kept the association with the largest 
number of counterparts. If the number of counterparts was the same,
we preferred the association having MIR data. If both associations (or 
none) had MIR data, we just kept the one with the highest probability.

Applying this method, we finally obtained a multi-wavelength catalogue
composed of 42\,705 X-ray sources with SDSS counterparts (u, g, r, i,
and z magnitudes), 14\,805 of them with near-infrared counterparts 
(UKIDSS or 2MASS; J, H, and K/K$_S$ magnitudes)\footnote{Although the 
wavelength response of K and K$_S$ filters is not exactly the same, we 
found that treating these magnitudes as equivalent did not significantly 
affect our results \citep[see][]{Mountrichas17}.}, and 26\,926 of them 
with WISE counterparts (W1 and W2 magnitudes). According to the 
distribution of association probabilities obtained in the ARCHES 
catalogue, we expect about 3660 miss-matches ($\sim 9\%$) in this 
catalogue.

\subsection{3XMM-Pan-STARRS (XPS)}
\label{subsec:xmatch}
Using data from PS1 we can increase the size of our final catalogue of 
photometric redshift, since it covers a larger sky area than the SDSS. 
Moreover, we can also include a significant fraction of the 3XMM sources 
not included in the ARCHES catalogue, either because they were rejected 
from the 3XMMe, or were new sources added in the 3XMM-DR6

We did our own cross-matching of PS1 and 3XMM-DR6 catalogues. 
Since we are interested in obtaining photometry in the NIR 
and MIR bands, we also included AllWISE, 2MASS, UKIDSS and 
VISTA-VHS surveys in the cross-match. We used the \texttt{xmatch} 
tool\footnote{\url{http://serendib.unistra.fr/ARCHESWebService/}} 
for this multi-catalogue cross-matching. This tool estimates the 
probability that a tuple of sources from different catalogues 
corresponds to the same real source (see \citealt{Pineau17} 
for a complete description of the algorithm). 

\texttt{xmatch} uses a Bayesian approach to estimate this probability, 
taking into account the likelihoods and priors of all possible 
hypotheses (e.g. all sources in the tuple are associated with different 
real sources, two sources in the tuple are associated with one real 
source and the rest to other source, etcetera). Likelihoods 
depend on the corresponding hypothesis, the number of cross-matched 
catalogues, and the Mahalanobis distance \citep{Mahalanobis36,DeMaesschalck00} 
between sources in the tuple, essentially an error weighted 
average of the positional distance. Priors, an estimation of the
probability that a particular hypothesis is obtained by chance, 
critically depend on the density of sources for each catalogue in 
the cross-matched area. Hence, to obtain meaningful probabilities 
the software has to be able to correctly estimate these densities.

\texttt{xmatch} cross-matches the catalogues using defined fields 
of common areas in all catalogues. These fields should be large 
enough to obtain an estimation of the source densities free of large 
statistical variance. In our case, the most straightforward way of 
defining the fields is through the XMM-\textit{Newton} observations 
included in the 3XMM. The XMM-\textit{Newton} field of view can be 
approximated by a circle of 15 arcmin radius. We selected all 3XMM 
observations included in the PS1 footprint (declination greater than 
-30~deg), except those flagged as OBS\_CLASS$>3$ (more than 10\% of 
the area of the observed field is identified as bad), or otherwise 
the compared areas between the 3XMM and the remaining catalogues would 
be significantly different. Since some of these observed fields have 
common areas, we grouped those overlapping fields in a single field. 
We end up with 2623 non-overlapping fields, covering a total of 
$\sim 630\;\mathrm{deg^2}$ ($\sim 466\;\mathrm{deg^2}$ outside the
Galactic plane). All PS1, 2MASS and AllWISE sources within these fields
were considered potential counterparts by \texttt{xmatch}, in addition 
to the corresponding UKIDSS-LAS and VISTA-VHS sources, if the field 
was included in those surveys.

\begin{figure}
  \centering
  \resizebox{\hsize}{!}{\includegraphics{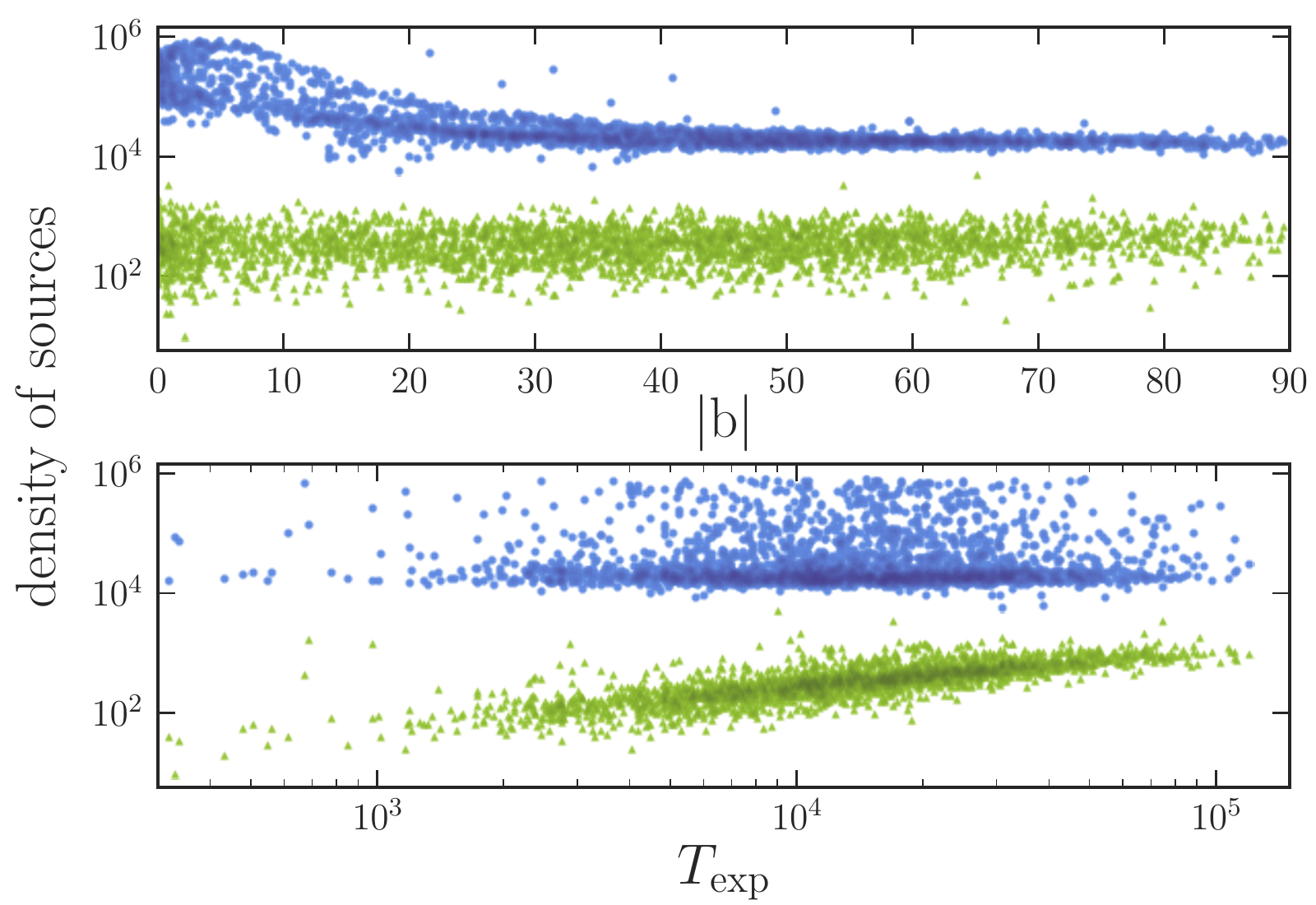}}
  \caption{Sky density of sources for the 2623 non-overlapping fields 
	we defined in the 3XMM/PS1 cross-match. Blue circles show the 
    density of optical sources, green triangles the density of X-ray 
    sources (darker shades show a higher density of data points).
	\textbf{Top:} Sky density versus the absolute value of the Galactic 
    latitude. 
	\textbf{Bottom:} Sky density versus exposure time of the 
    XMM-\textit{Newton} observation.}
\label{fig:skydensity}
\end{figure}

However, the areas of most of these fields are not large enough to 
obtain a good estimation of the density of X-ray sources. This density 
depends mostly on the exposure time of the corresponding XMM-\textit{Newton} 
observation (see Fig.~\ref{fig:skydensity}, bottom).\footnote{We estimated 
the exposure time of the observation as a weighted average (weights 
3, 1, 1) of the exposure time of the three EPIC cameras on-board 
XMM-\textit{Newton} (PN, MOS1, MOS2)}. By grouping fields of similar 
exposure time we can reduce the statistical variance and get a better 
estimation of the density of the X-ray sources. We must also take 
into account that the density of PS1 sources depends on the Galactic 
latitude (Fig.~\ref{fig:skydensity}, top). Based on these two 
parameters, we grouped the 2623 fields in bins with roughly constant 
densities of X-ray and optical sources. Using this method, we defined 
70 different bins, with sky areas ranging between $\sim 4-20\;\mathrm{deg^2}$, 
and we run \texttt{xmatch} for each one.

Given the high density of optical sources in the Galactic plane, the 
association probabilities calculated by \texttt{xmatch} in this area 
are extremely low. A large difference in the density of sources of 
two catalogues implies that the probability of association 
by chance is very high. In the Galactic plane the difference between 
the density of optical and X-ray sources can reach up to four orders 
of magnitude (see Fig.~\ref{fig:skydensity}). We therefore decided to 
reject those sources from our catalogue and to keep only objects outside 
the Galactic plane (i.e. $|b|>20$~deg).

For consistency with the ARCHES data set, for building our final 
catalogue we followed the same photometry- and probability-based 
selection criteria we explained in Sect.~\ref{subsec:arches} above. The 
only difference is that we did not marginalise the probabilities of 
association, because our cross-match did not contain catalogues we 
were not interested in. Therefore, the \texttt{xmatch} output directly 
gave us the proper probabilities.

Our final 3XMM/Pan-STARRS catalogue (XPS) is composed of 88\,088 X-ray 
sources with PS1 counterparts (g, r, i, z, and y magnitudes), 21\,174 of 
them with NIR counterparts (2MASS, UKIDSS or VISTA-VHS, J, H, and K or K$_S$ 
magnitudes), and 54\,947 of them with MIR (All-WISE, W1 and W2 magnitudes) 
counterparts. According to the distribution of association probabilities 
given by \texttt{xmatch}, we expect about 8100 miss-matches ($\sim 9\%$) 
in this catalogue.

\section{Photometric redshifts}
\label{sec:photoz}
We used a supervised machine-learning technique to estimate photometric 
redshifts for XMMPZCAT. These techniques require a spectroscopic sample 
to train an algorithm (Sect.~\ref{subsec:training}). Then the algorithm 
is applied to a dataset with available photometry to estimate photometric 
redshifts. 

Our photo-z determination is primary based on the five optical bands 
available in SDSS and PS1: g, r, i, z, plus u or y respectively, 
depending on the optical catalogue we are using. We tested that
including photometric points beyond the optical range significantly 
improves the accuracy of photo-z \citep[see also e.g.][]{Ball07,Rowan08,Yang17}, 
and decrease the number of outliers. Hence, whenever available we 
have also used NIR and/or MIR photometry to derive the photometric 
redshifts. 

Photometric redshifts were estimated using 
MLZ-TPZ\footnote{\url{http://matias-ck.com/mlz/}} \citep{tpz}, a 
machine-learning algorithm based on a supervised technique with 
prediction trees and random forest. It is a parallelizable python 
package that calculates fast and robust photometric redshifts and 
their corresponding probability density functions (PDF).

Prediction trees are a non-linear technique for solving classification 
or regression problems \citep{Breiman84}. The trees are build recursively
splitting the training sample using a set of criteria based on the 
properties of the sample (e.g. magnitudes), until a defined stop condition
is reached. These criteria are estimated in a way that maximises, for 
each split, the informational gain in the parameter of interest (e.g. 
redshift). TPZ implements two types of trees: classification and regression 
trees, depending if the redshift is considered a discrete or a continuous 
parameter. We used regression trees for XMMPZCAT.

A random forest algorithm \citep{Breiman01} first generates several 
predictions trees and then combines all possibles outcomes to give a 
final prediction. Random forests are one of the most robust and accurate 
supervised learning techniques available today \citep{Caruana08}.

TPZ generates $N_R$ training samples by perturbing the input properties of
the original training sample, according to the errors of each variable and
assuming that they are normally distributed. For each training sample, a 
prediction tree is generated. Then TPZ, by applying bootstraping, 
creates $N_T$ new trees for each of those previously generated trees. 
Finally, all trees are combined in a single random forest, containing 
$N_R \times N_T$ trees.

\begin{table}[t]
  \caption{Selected colours for the estimation of photo-z.}
  \label{tab:colourset}
  \centering
  \begin{tabular}{ll}
  \multicolumn{2}{c}{SDDS} \\
  \hline
  Sample & Colours \\
  \hline
  \hline
    10 filters & u$-$g, g$-$r, g$-$i, g$-$z, r$-$i, r$-$z, i$-$z, z$-$W1, W1$-$W2, \\
               & K$-$W1, J$-$W1, H$-$W1, J$-$H, H$-$K, J$-$K \\
     8 filters & u$-$g, g$-$r, g$-$i, g$-$z, r$-$i, r$-$z, i$-$z, z$-$J, J$-$H \\
     7 filters & u$-$g, g$-$r, g$-$i, g$-$z, r$-$i, r$-$z, i$-$z, z$-$W1, W1$-$W2 \\
     5 filters & u$-$g, g$-$r, g$-$i, g$-$z, r$-$i, r$-$z, i$-$z \\
  & \\
  \multicolumn{2}{c}{Pan-STARRS} \\
  \hline
  Sample & Colours \\
  \hline
  \hline
    10 filters & g$-$r, g$-$i, g$-$z, r$-$i, r$-$z, i$-$z, i$-$y, z$-$W1, W1$-$W2, \\
               & K$-$W1, J$-$W1, H$-$W1, J$-$H, H$-$K, J$-$K \\
     8 filters & g$-$r, g$-$i, g$-$z, r$-$i, r$-$z, i$-$z, i$-$y, z$-$J, J$-$H \\
     7 filters & g$-$r, g$-$i, g$-$z, r$-$i, r$-$z, i$-$z, i$-$y, z$-$W1, W1$-$W2 \\
     5 filters & g$-$r, g$-$i, g$-$z, r$-$i, r$-$z, i$-$z, i$-$y \\
  \hline
\end{tabular}
\end{table}

An additional source of randomness can be added if each one of the $N_T$ 
trees is generated using only a limited subset of $m_*$ properties from 
the total set of input properties (e.g. if there are 8 magnitudes available 
in the training sample, select only 5 to generate a particular forest). 
In this work we used $N_R=15, N_T=10, m_*=4$. 

With these parameters we can obtain a random forest of 150 trees. As shown 
in \citet[Fig.~9]{tpz}, using a larger forest does not significantly increase 
the predictive power of this technique. Even more, given the size of our 
training sample, we found that using a forest with more than about 50 trees 
did not significantly change our photo-z estimates \citep{Mountrichas17}.
However, increasing the number of trees up to 150 does in fact improve the 
estimate of the PDF.

The photo-z for the application sample are then calculated using all these 
prediction trees. Each source in the sample runs down each tree making a 
prediction. Combining the predictions of all trees the corresponding PDF 
is generated. The photo-z included in XMMPZCAT is the most probable value 
(i.e. the mode of the PDF).

We employed colours instead of magnitudes as input data for TPZ. Colours 
were corrected of Galactic extinction using the \citet{Schlafly11} 
extinction maps, as given by the Galactic Dust Reddening and Extinction 
Service.\footnote{\url{http://irsa.ipac.caltech.edu/applications/DUST/}} 
Using colours we minimise the effect of the training sample being brighter 
in the optical than the actual data for which we have to predict the
redshifts (see Sect.~\ref{subsec:training}).

We divided our training and application samples into point-like and extended 
sources, and according to the photometric data available (see 
Sect.~\ref{subsec:training} below). Table~\ref{tab:colourset} shows
the selected colour set for each training sample. They were selected 
after extensive testing to optimise our photo-z estimations, using the 
relative importance given by TPZ to each colour \citep[see][]{tpz}.

We derived photometric redshifts for 42\,705 X-ray sources in the ARCHES 
catalogue and for 88\,088 sources in the XPS catalogue. Taking into 
account the number of sources that both catalogues share in common 
($\sim 30\,000$, see Sect.~\ref{subsec:merge}), we end up with a catalogue 
of photometric redshifts for 100\,178 objects. The photo-z of 20\,025 
sources (20\%) were estimated using photometry in ten filters (optical, 
NIR and MIR), 6929 sources (7\%) with photometry in eight filters (optical 
and NIR),41\,973 (42\%) with seven filters (optical and MIR) and 31\,251 
(31\%) with only five filters (optical).

The XMMPZCAT can be downloaded as a fits table in the web site of
this project.\footnote{\url{http://xraygroup.astro.noa.gr/Webpage-prodex/xmmpzcat_access.html}}
We also provide an auxiliary table containing the photo-z PDF of 
each source. We give a detailed description of the catalogue in 
the Appendix. 

\begin{figure}
  \centering
  \resizebox{\hsize}{!}{\includegraphics{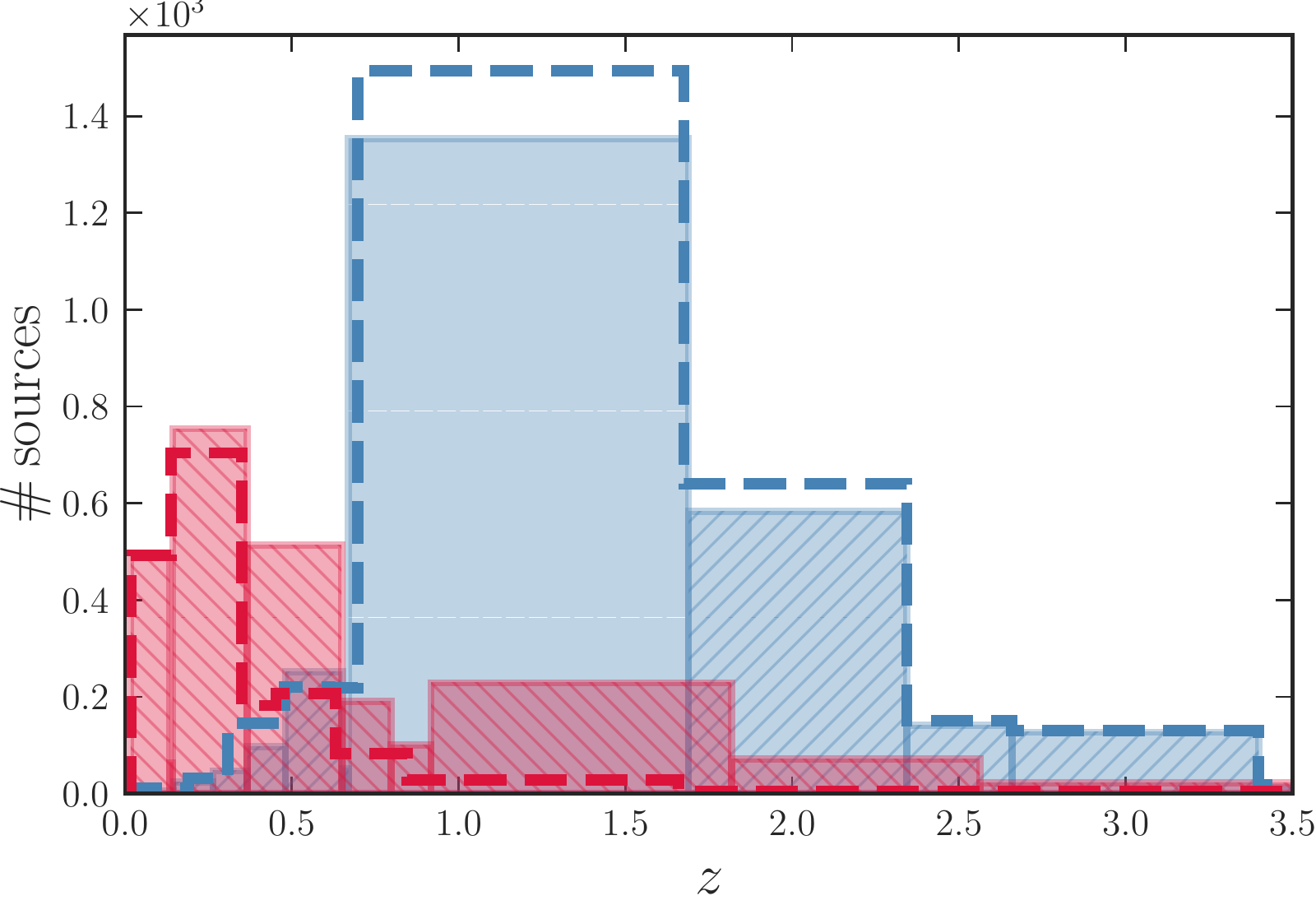}}
  \caption{Redshift distribution of our training samples (five filters). 
  	Solid histograms correspond to the SDSS samples, and open, dashed 
    histograms to the Pan-STARRS samples. Extended (point-like) sources 
    are represented in red (blue). The histograms' binning was estimated 
    using the Bayesian block algorithm \citep[BBA;][]{Scargle13}.}
\label{fig:zdist_training}
\end{figure}

\begin{figure*}
  \centering
  \resizebox{\hsize}{!}{\includegraphics{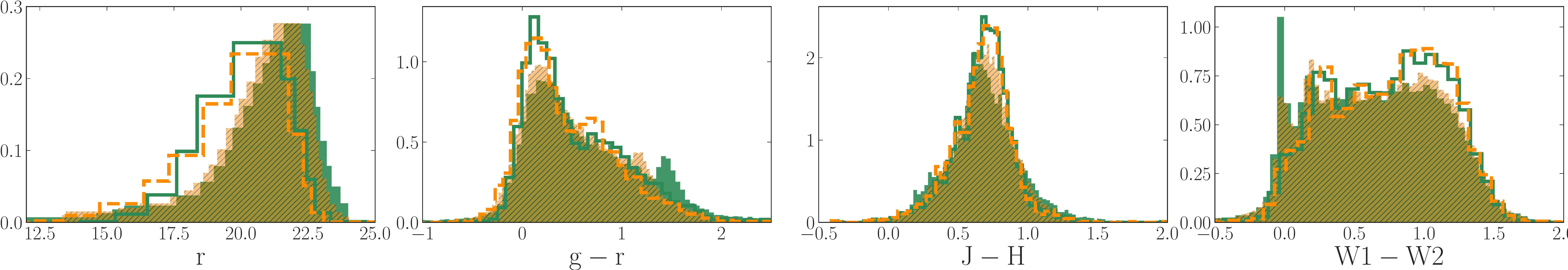}}
  \caption{Normalised distributions of r magnitude and g$-$r, J$-$H and 
  W1$-$W2 colours for our training samples and the corresponding application 
  samples. SDSS training sample: green, open histogram; Pan-STARRS training 
  sample: yellow, dashed, open histogram; ARCHES sample: green, solid 
  histogram; XPS sample: yellow, hatched histogram.}
\label{fig:coldist}
\end{figure*}

\begin{table*}[t]
 \caption{Training samples: size and redshift distribution statistics.}
 \label{tab:training}
 \centering
 \begin{tabular}{ccccc||cccc}
  \hline
          & \multicolumn{4}{c||}{SDDS} & \multicolumn{4}{c}{Pan-STARRS} \\  
  \hline    
  \# filters & \# sources & $z_\mathrm{Q10}$\tablefootmark{a} & 
                          $z_\mathrm{median}$\tablefootmark{b} & 
                          $z_\mathrm{Q90}$\tablefootmark{c} &
             \# sources & $z_\mathrm{Q10}$\tablefootmark{a} & 
                          $z_\mathrm{median}$\tablefootmark{b} & 
                          $z_\mathrm{Q90}$\tablefootmark{c} \\
  \hline
   & \multicolumn{8}{c}{Point-like sources} \\
  \hline
  \hline
  5  & 2703 & 0.55 & 1.30 & 2.38 & 3063 & 0.55 & 1.26 & 2.34 \\
  7  & 2420 & 0.55 & 1.29 & 2.34 & 2726 & 0.54 & 1.23 & 2.31 \\
  8  & 1508 & 0.59 & 1.35 & 2.43 & 1687 & 0.57 & 1.25 & 2.39 \\
  10 & 1429 & 0.59 & 1.34 & 2.42 & 1593 & 0.57 & 1.24 & 2.36 \\
  \hline
   & \multicolumn{8}{c}{Extended sources} \\
  \hline
  \hline
  5  & 2454 & 0.07 & 0.34 & 1.16 & 1710 & 0.06 & 0.23 & 0.57 \\
  7  & 2298 & 0.07 & 0.32 & 0.95 & 1670 & 0.06 & 0.23 & 0.57 \\
  8  & 1621 & 0.06 & 0.25 & 0.77 & 1294 & 0.06 & 0.20 & 0.53 \\
  10 & 1586 & 0.06 & 0.25 & 0.75 & 1283 & 0.06 & 0.20 & 0.52 \\

  \hline
 \end{tabular}
\tablefoot{
\tablefoottext{a}{Tenth percentile of the redshift distribution.}
\tablefoottext{b}{Median of the redshift distribution.}
\tablefoottext{c}{Ninetieth percentile of the redshift distribution.}
}
\end{table*}

\subsection{Training samples}
\label{subsec:training}
One of the key aspects of estimating photometric redshifts through 
supervised machine learning methods is the selection of an adequate 
training sample. This sample should be representative of the global 
sample for which the photo-z will be calculated. In our case we need 
two training samples, one for the ARCHES catalogue and one for our 
XPS catalogue.

We used the training sample presented in \citet{Mountrichas17} for
the sources in the ARCHES catalogue (SDSS training sample). This sample 
contains sources from XXL \citep{Menzel16}, XWAS \citep{xwas}, COSMOS 
\citep{cosmosxmm}, XMS \citep{xms} and XBS \citep{DellaCeca04}, all of 
them X-rays surveys with a high level of spectroscopic identification. 
In addition, it also contains 1500 SDSS-DR13 sources spectroscopically 
identified as QSO with X-ray counterparts. Even though these QSO are 
optically selected instead of X-ray selected, adding them does not bias 
our photo-z derivation significantly \citep{Mountrichas17}. The final 
training sample is composed of 5157 objects with SDSS photometric 
data, 3129 with also NIR data (UKIDSS or 2MASS) and 4718 with MIR 
data (AllWISE). 

This sample was also our starting point for building a training sample 
for our XPS catalogue. We did a positional cross-match between the SDSS 
training sample and PS1. We selected matches with an angular separation 
$<0.3$ arcsec and with good photometry, that is sources with a photometric 
measurement and the corresponding error (no upper limits included) in all five 
filters (g, r, i, z, y). The resulting sample contains 4773 objects with 
spectroscopic redshifts and PS1 photometry. 2981 of them have also NIR 
data, and 4396 have MIR data.

Both training samples were split in eight different subsets by dividing 
each sample according to the sources extension in the optical (whether 
the source is classified as a point-like or extended object in the 
corresponding optical survey), and the amount of photometric data 
available: only optical (five filters: [u]griz[y]), optical+NIR (eight 
filters: [u]griz[y]JHK), optical+MIR (seven filters: [u]griz[y]W1W2), 
and optical+NIR+MIR (ten filters: [u]griz[y]JHKW1W2). Table~\ref{tab:training} 
shows the size and statistics for the redshift distributions of all the 
training samples.

\begin{figure}
  \centering
  \resizebox{\hsize}{!}{\includegraphics{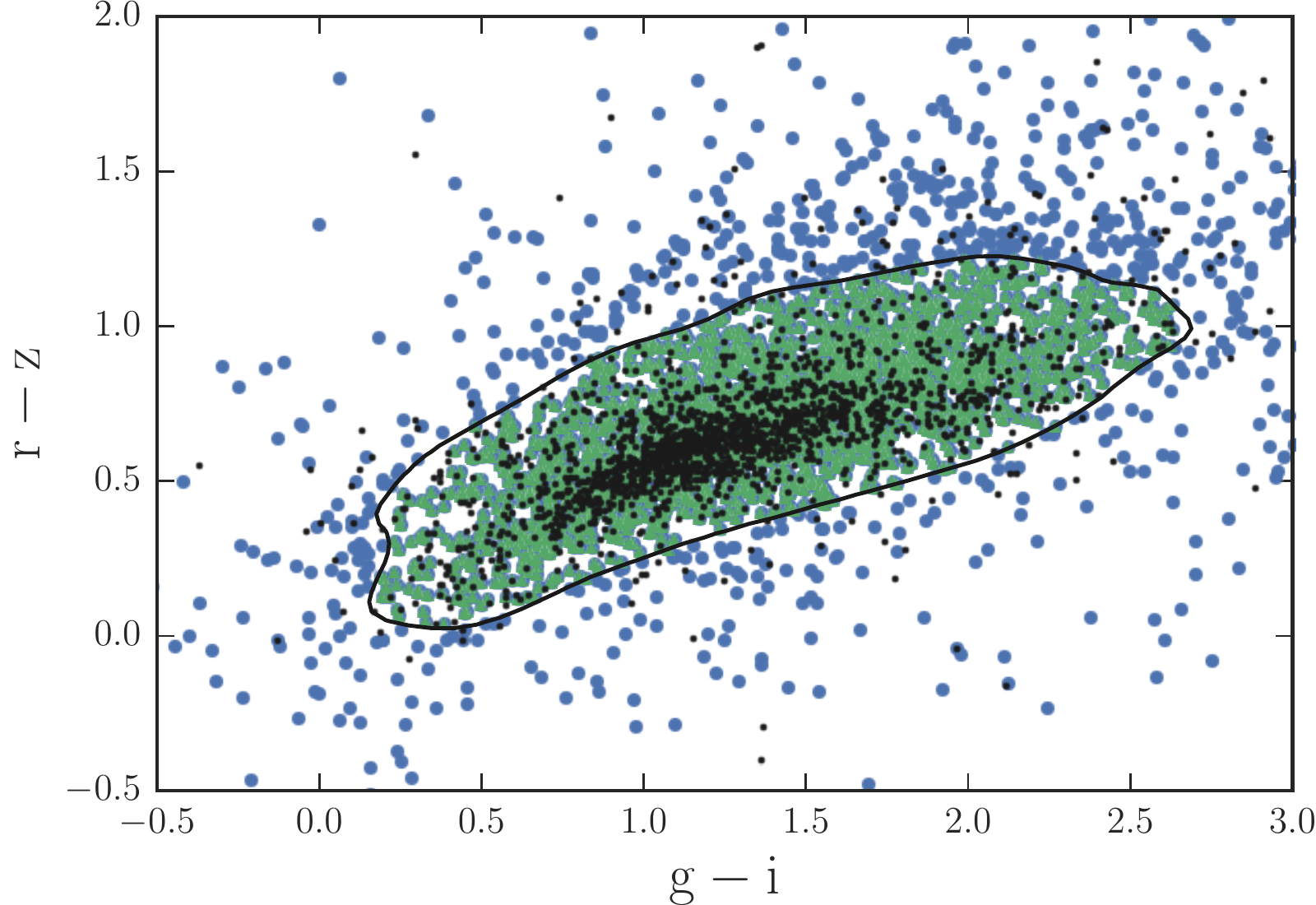}}
  \caption{g$-$r against r$-$z for extended sources with only optical counterparts 
  in the XPS sample. Black points represent the corresponding training sample. The 
  black solid line shows the 90\% limit of the kernel density estimation for the 
  training sample (i.e. the colour space region well covered by the training sample. 
  Green triangles are sources inside the 90\% limit, blue circles sources outside 
  the 90\% limit.}
\label{fig:gmirmz_training}
\end{figure}

\begin{figure*}
  \centering
  \resizebox{\hsize}{!}{\includegraphics{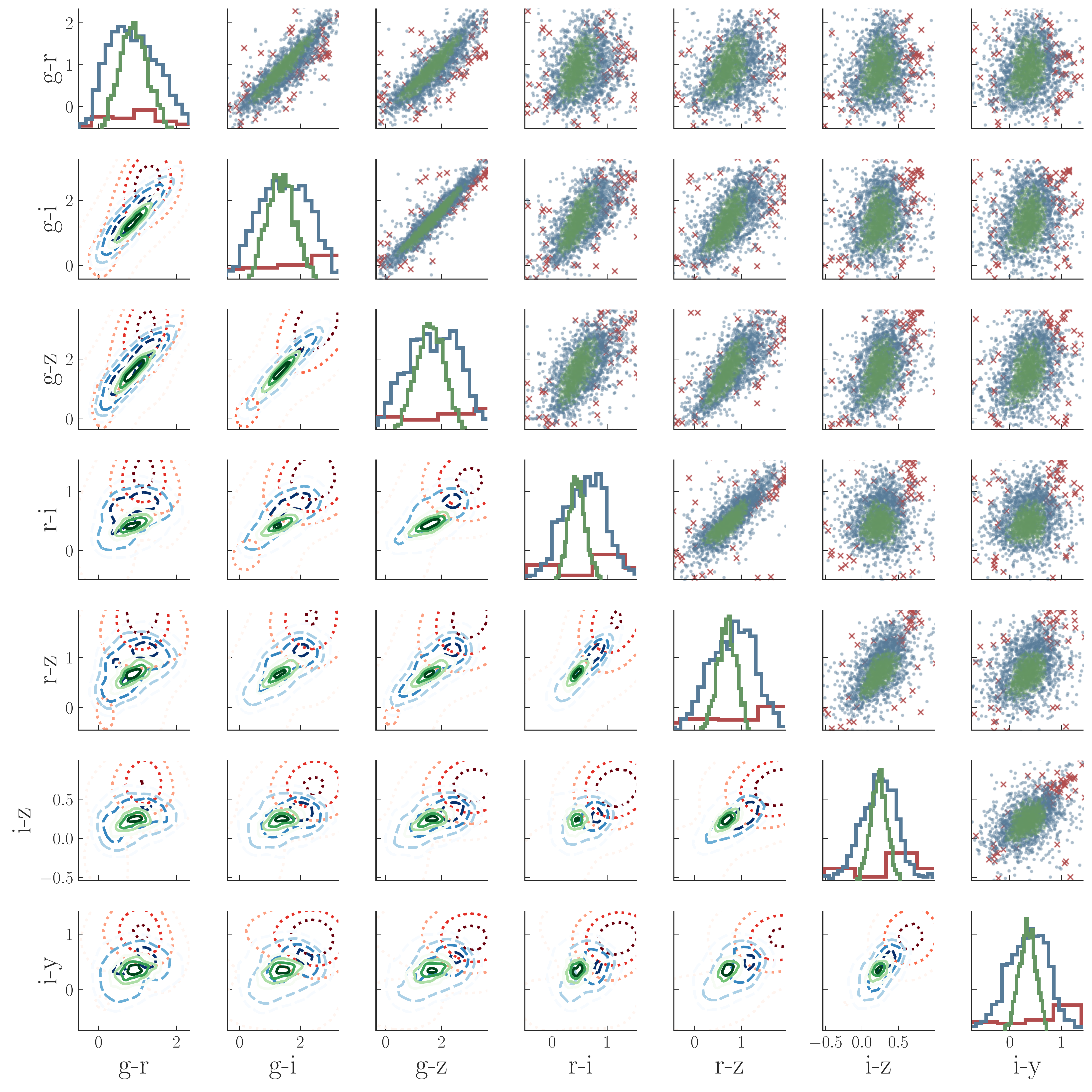}}
  \caption{\textbf{Diagonal:} Optical colour distributions for extended sources 
  with only optical counterparts in the XPS catalogue. 
  \textbf{Above diagonal:} colour-colour plots. 
  \textbf{Below diagonal:} kernel density estimations of the colour distribution.  
  Green (triangles in the colour-colour plots, solid lines in the density contour 
  plots): sources with all colours within the well covered region of the training 
  sample (\texttt{inTSCS} is true, see Appendix); blue (circles in the colour-colour 
  plots, dashed lines in the density contour plots): sources with at least one colour 
  within the well covered region of the training sample; red (crosses in the 
  colour-colour plots, dotted lines in the density contour plots): sources with all 
  colours outside the well covered region of the training sample.}
\label{fig:colcol_training}
\end{figure*}

Figure~\ref{fig:zdist_training} shows the redshift distributions
for the SDSS (solid histogram) and Pan-STARRS (open, dashed histogram) 
five filters training samples, divided in extended (red) and point-like 
(blue) sources. The overall distribution of both samples is quite similar,
with the majority of sources below $z \sim 3$. The main difference
is in how extended and point-like objects are distributed. Most extended
sources of the Pan-STARRS training sample show $z \lesssim 1$, while we find 
a non-negligible number of extended sources in the SDSS training sample
between $1 \lesssim z \lesssim 2$. 

Figure~\ref{fig:coldist} presents the r magnitude, g$-$r, J$-$H and W1$-$W2 
colour distributions of the training samples (open histograms) and the 
ARCHES and XPS full catalogues (solid histograms). Although our training 
samples are about one magnitude brighter than the application samples, 
their colour distributions are reasonably well reproduced. As noted by 
\citet{Beck17}, this is an important factor to obtain reliable photometric 
redshifts. The only major differences are the peaks at $\mathrm{g-r} \sim 1.5$ 
and $\mathrm{W1-W2} \sim 0$, which are caused by the underlying population 
of stars in our application samples (see Sect.~\ref{subsec:stars}).

For a more quantitative estimation of how well our application samples are
covered by their corresponding training samples, we estimated the parameter
\texttt{inTSCS} (in Training Sample Colour Space, see Appendix)
for each source in our final catalogue: For each possible independent combination
of two colours (e.g. g$-$r against z$-$y, W1$-$W2 against u$-$g, etcetera) we 
did a kernel density estimation (KDE) of the distribution of training sample 
sources in that colour space, and we estimated the 90\% probability contour
of the KDE (see e.g. Fig.~\ref{fig:gmirmz_training}). Sources of the application
sample inside this contour are well covered by the training sample (green 
triangles in Fig.~\ref{fig:gmirmz_training}). Sources well covered in all 
possible colour combinations (see Fig.~\ref{fig:colcol_training}) have a 
true \texttt{inTSCS} value in XMMPZCAT. About $\sim 40-50\%$ of the sources 
are well covered in all colour combinations, and $\sim 85-95\%$ are well 
covered in at least one colour-colour combination, depending on the particular training sample. Considering the whole XMMPZCAT, 45\% of the sources have a 
true \texttt{inTSCS} value.

\begin{figure}
  \centering
  \resizebox{\hsize}{!}{\includegraphics{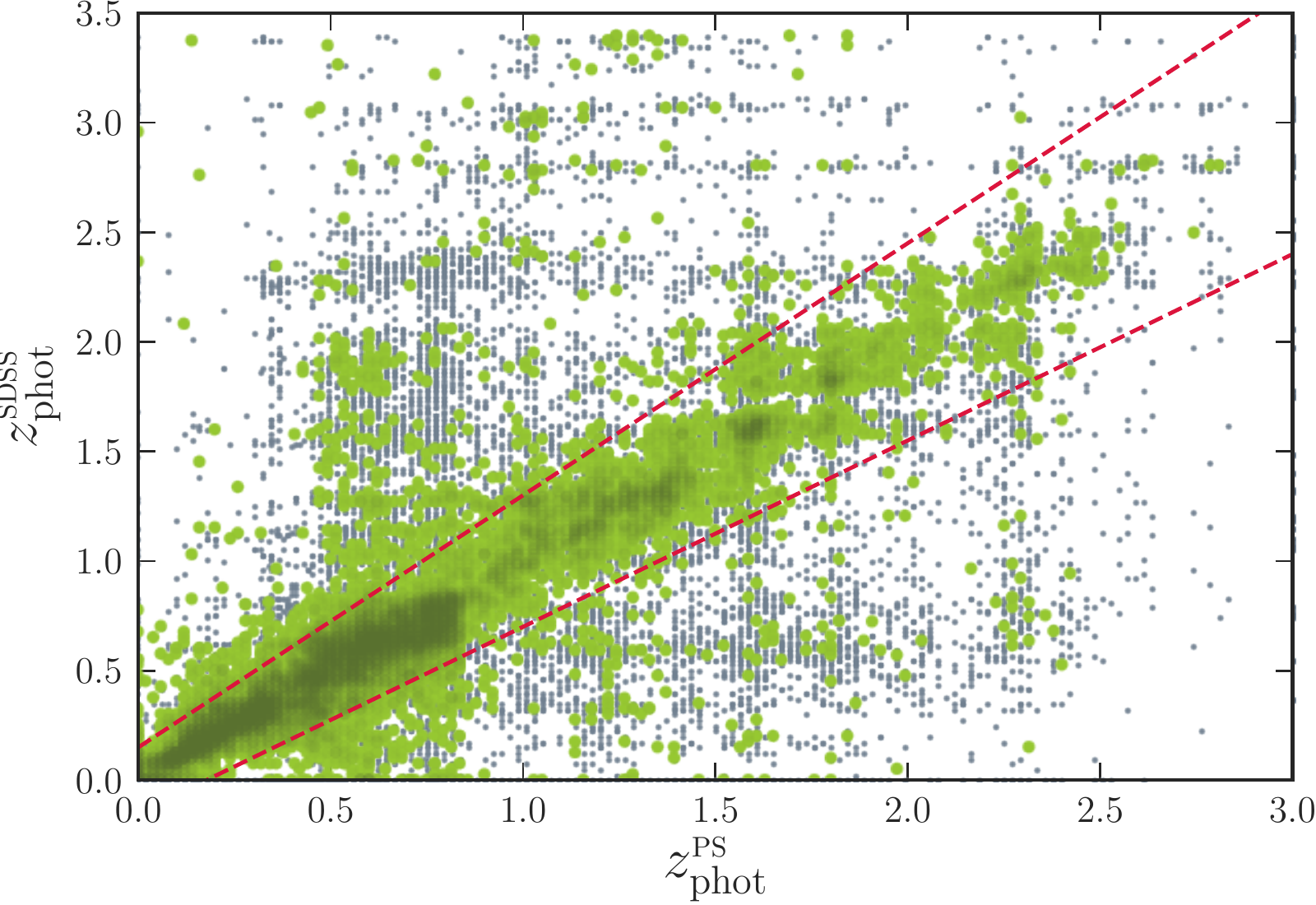}}
  \caption{Comparison of photometric redshifts for 3XMM sources with 
  	optical counterparts in both SDSS and Pan-STARRS catalogues. Green
	circles are objects having both photo-z with PS greater than 0.7 
    (see Sect.~\ref{subsec:zquality}). Darker shades of green show 
    higher density of sources. Grey points are the remaining sources. 
    Red, dashed lines show the limit where 
    $|z_\mathrm{phot}^\mathrm{\scriptscriptstyle PS}-
    z_\mathrm{phot}^\mathrm{\scriptscriptstyle SDSS}|/(1-
    z_\mathrm{phot}^\mathrm{\scriptscriptstyle PS})=0.15$.}
\label{fig:zphots_common}
\end{figure}

\subsection{Merging ARCHES and XPS results}
\label{subsec:merge}
We found 32\,460 X-ray sources with counterparts in both ARCHES and XPS
catalogues. In order to build our final catalogue of photometric redshifts, we 
have to select which photo-z include for these sources. Our merging criterion 
was as follows: in those cases where the ARCHES photo-z was estimated using 
the same or more filters than the XPS photo-z, and the SDSS photometry was 
flagged as clean (see Appendix), we selected the ARCHES 
photo-z; otherwise the XPS photo-z was included in the final catalogue.

We can use these common sources as a consistency check of our results. If 
the SDSS and Pan-STARRS counterparts of the X-ray source are in fact the 
same optical source (i.e. disregarding miss-matching problems), we expect 
that the photo-z estimated either with SDSS or Pan-STARR photometry should 
be similar. Figure~\ref{fig:zphots_common} shows a comparison between the 
ARCHES and XPS photo-z of these 32\,460 common objects. We define that a 
source has consistent photo-z if 
$|z_\mathrm{phot}^\mathrm{\scriptscriptstyle PS}-z_\mathrm{phot}^\mathrm{\scriptscriptstyle SDSS}|/(1-z_\mathrm{phot}^\mathrm{\scriptscriptstyle PS})<=0.15$.
Using this definition, about 70\% of the common sources show consistent 
photo-z. Restricting the sample to sources having reliable photo-z in 
both catalogues (i.e. with peak strength equal or greater than 0.7, see 
Sect.~\ref{subsec:zquality}), the fraction of sources with consistent 
photo-z is 90\% (green circles in Fig.~\ref{fig:zphots_common}).

\begin{figure*}
  \centering
  \resizebox{\hsize}{!}{\includegraphics{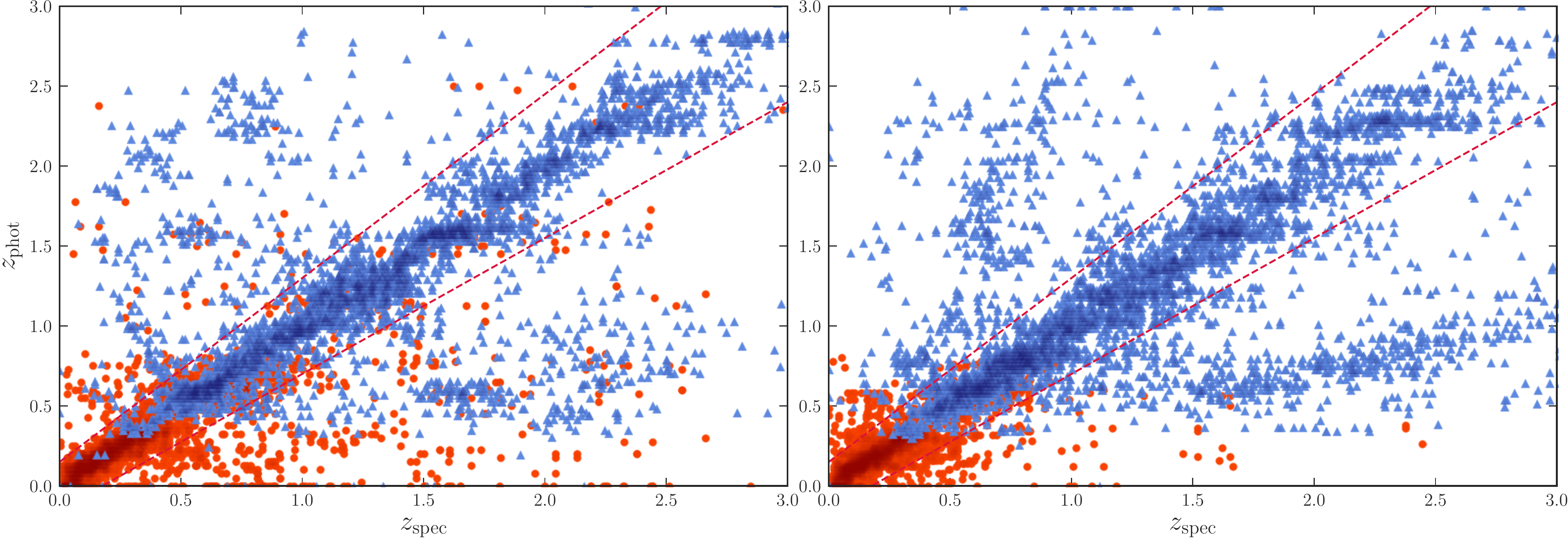}}
  \caption{Spectroscopic against photometric redshifts of SDSS (left) and 
           Pan-STARRS (right) training samples. Blue triangles are 
           point-like sources, red circles are extended sources. Darker shades
           represent a higher density of sources. Red, dashed lines show 
           the adopted limit for catastrophic outliers (see 
           Sect.~\ref{subsec:zquality}).}
\label{fig:zphotzspec}
\end{figure*}

\section{Results}
\label{sec:results}

\begin{table*}
  \caption{Results of the statistical tests using the training+testing 
  samples (Sect.~\ref{subsec:zquality}).}
  \label{tab:test1}
  \centering
\begin{tabular}{ccc|cc|cc||cc|cc|cc}
  \hline
          & \multicolumn{2}{c}{(1)} & 
            \multicolumn{2}{c}{(2)} & 
            \multicolumn{2}{c}{(3)} & 
            \multicolumn{2}{c}{(4)} &
            \multicolumn{2}{c}{(5)} &
            \multicolumn{2}{c}{(6)} \\              
          & \multicolumn{2}{c}{SDDS} & 
            \multicolumn{2}{c}{SDDS-b\tablefootmark{a}} & 
            \multicolumn{2}{c||}{SDDS-c\tablefootmark{c}} & 
            \multicolumn{2}{c}{Pan-STARRS} &
            \multicolumn{2}{c}{Pan-STARRS-b \tablefootmark{b}} &
            \multicolumn{2}{c}{Pan-STARRS-c \tablefootmark{c}} \\  
  \hline    
  \# filters\tablefootmark{d} & 
  $\sigma_\mathrm{NMAD}$ & $\eta$ (\%) & 
  $\sigma_\mathrm{NMAD}$ & $\eta$ (\%) & 
  $\sigma_\mathrm{NMAD}$ & $\eta$ (\%) & 
  $\sigma_\mathrm{NMAD}$ & $\eta$ (\%) & 
  $\sigma_\mathrm{NMAD}$ & $\eta$ (\%) &   
  $\sigma_\mathrm{NMAD}$ & $\eta$ (\%) \\
  \hline
  \multicolumn{13}{c}{Point-like sources} \\
  \hline
  \hline
   5 (4) & 0.076 & 29 & 0.089 & 32 & 0.132 & 39 & 0.138 & 41 & 0.135 & 41 & 0.168 & 45 \\
   7 (6) & 0.064 & 19 & 0.072 & 22 & 0.082 & 22 & 0.088 & 25 & 0.087 & 27 & 0.096 & 27 \\
   8 (7) & 0.057 & 20 & 0.059 & 21 & 0.069 & 24 & 0.074 & 26 & 0.067 & 28 & 0.087 & 28 \\
  10 (9) & 0.049 & 14 & 0.049 & 13 & 0.059 & 20 & 0.062 & 17 & 0.069 & 21 & 0.069 & 17 \\
  \hline
  \multicolumn{13}{c}{Extended sources} \\
  \hline
  \hline
   5 (4) & 0.071 & 18 & 0.061 & 11 & 0.091 & 28 & 0.063 & 13 & 0.078 & 21 & 0.072 & 19 \\
   7 (6) & 0.057 & 14 & 0.048 &  8 & 0.057 & 19 & 0.038 &  6 & 0.047 & 11 & 0.039 &  5 \\
   8 (7) & 0.054 & 12 & 0.051 &  7 & 0.063 & 13 & 0.052 &  9 & 0.058 & 10 & 0.056 &  9 \\
  10 (9) & 0.046 &  9 & 0.043 &  6 & 0.051 & 10 & 0.036 &  4 & 0.041 &  7 & 0.037 &  4 \\
  \hline
\end{tabular}
\tablefoot{
\tablefoottext{a}{Test using only sources with $z<1$ in the extended samples.}
\tablefoottext{b}{Test with the extended/point-like splitting using only the 
                  SDSS classification.}
\tablefoottext{c}{Test using only optical data from the common filters between the 
                  SDSS and Pan-STARRS samples (g, r, i, z)}
\tablefoottext{d}{The numbers inside the parenthesis refers to the number of filters 
                  in the SDSS-c and Pan-STARRS-c samples.}
}
\end{table*}

\begin{figure}
  \centering
  \resizebox{\hsize}{!}{\includegraphics{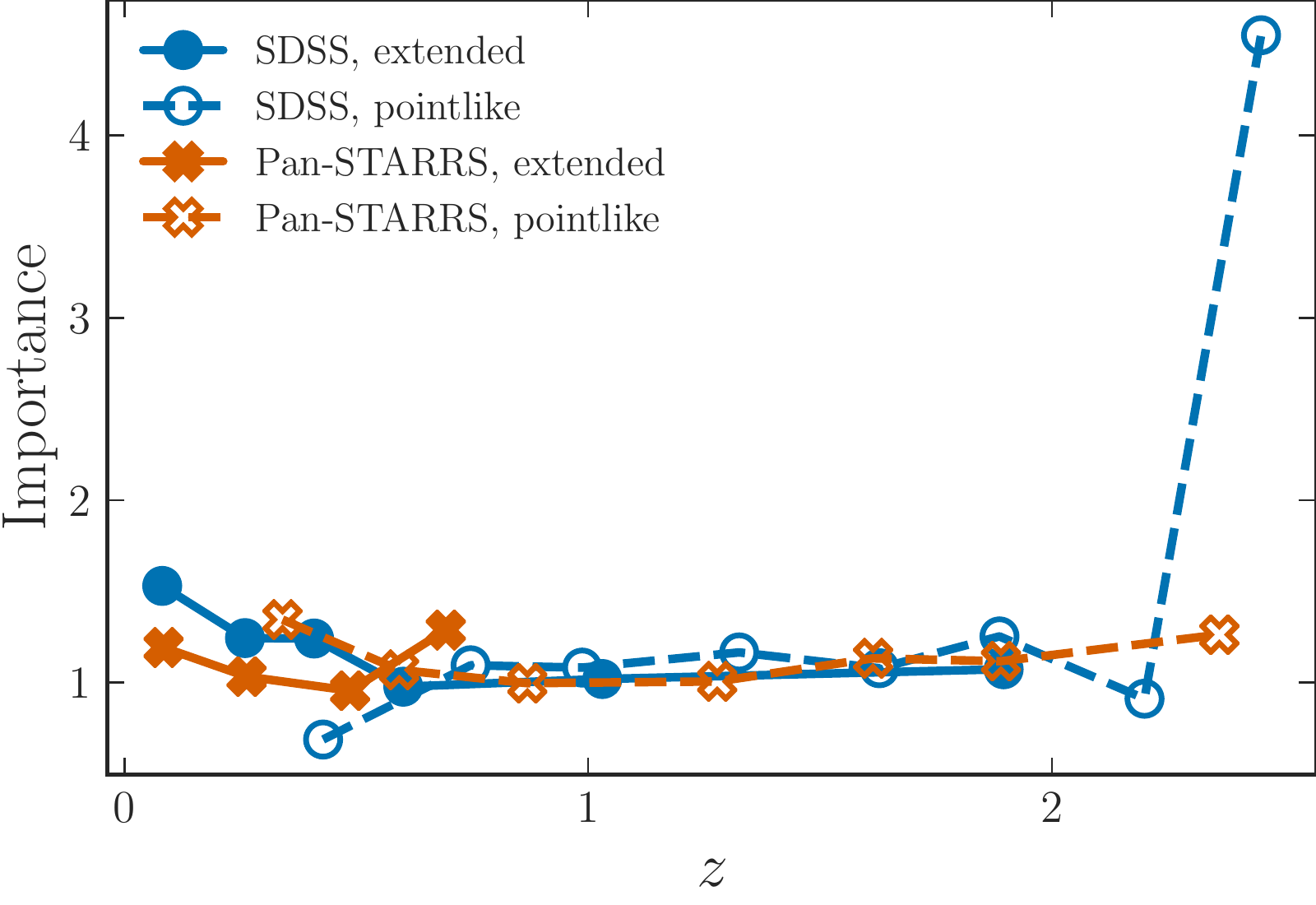}}
  \caption{Relative importance of u$-$g (blue circles) and i$-$y 
  (orange crosses) against redshift for the SDSS and Pan-STARRS 5 filters 
  training samples, respectively (see Sect.~\ref{subsec:zquality}). Solid 
  lines and symbols: extended subsamples; open symbols and dashed lines: 
  point-like subsamples.}
\label{fig:uyimportance}
\end{figure}

\begin{figure*}
  \centering
  \resizebox{\hsize}{!}{\includegraphics{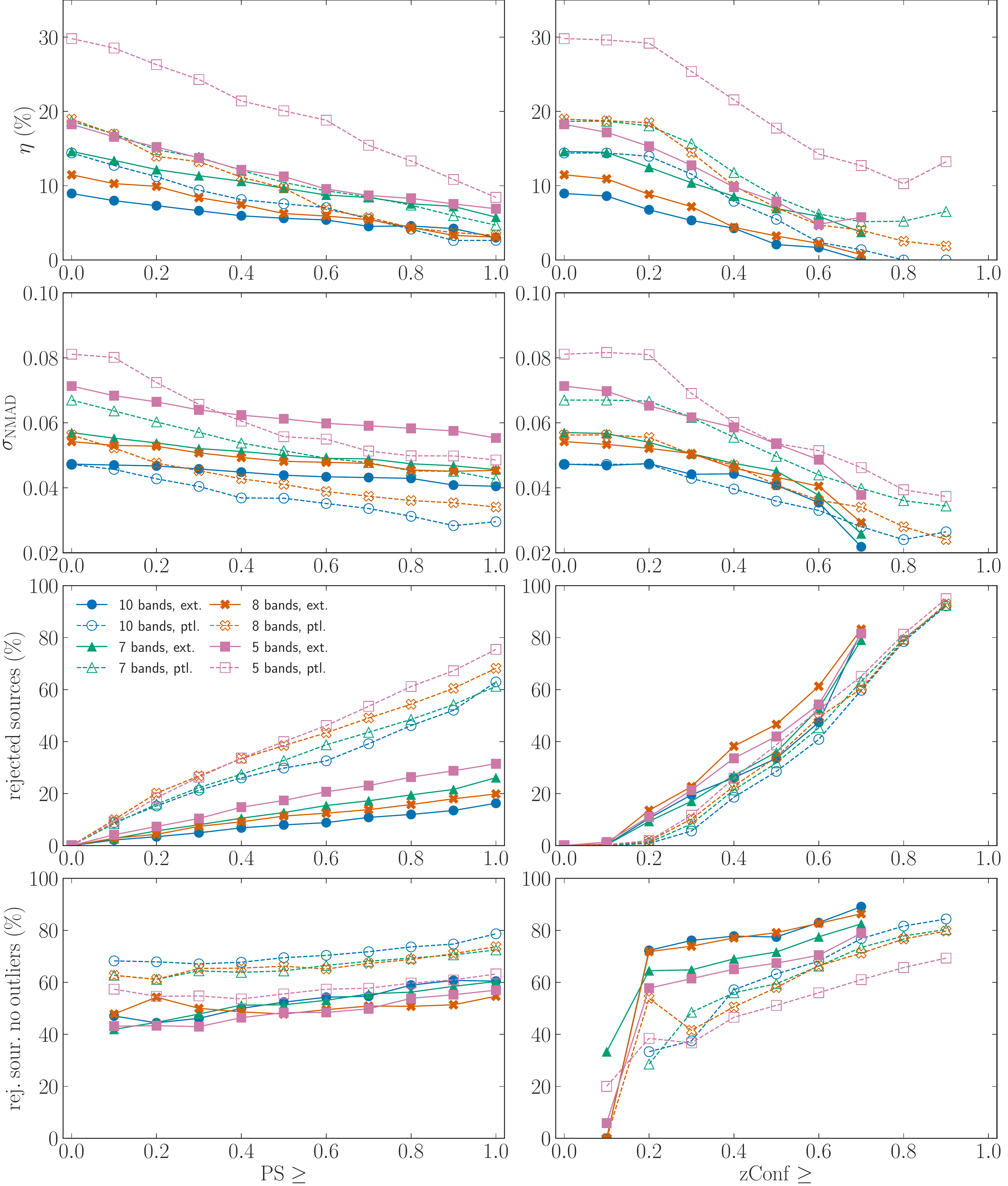}}
  \caption{Statistics for the derivation of photometric redshifts, using 
  	the training+testing SDSS samples. Left column plots correspond to 
  	a PS-based selection, right column plots correspond to a zConf selection 
  	(see Sect.~\ref{subsec:zquality}). \textbf{First row:} percentage of 
    outliers; \textbf{second row:} normalised median absolute deviation; 
    \textbf{third row:} percentage of rejected sources after filtering; 
    \textbf{fourth row:} percentage of rejected sources that are not outliers. 
    Blue circles correspond to ten filters samples (ugrizJHKW1W2), green 
    traingles to seven filters samples (ugrizW1W2), orange crosses to eight 
    filters samples (ugrizJHK), purple squares to five filters samples (ugriz). 
    Solid lines and symbols correspond to extended sources samples, dashed 
    lines and empty symbols to point-like sources samples. Lines are included 
    just for visualisation improvement.}
\label{fig:zquality_arches}
\end{figure*}

\begin{figure*}
  \centering
  \resizebox{\hsize}{!}{\includegraphics{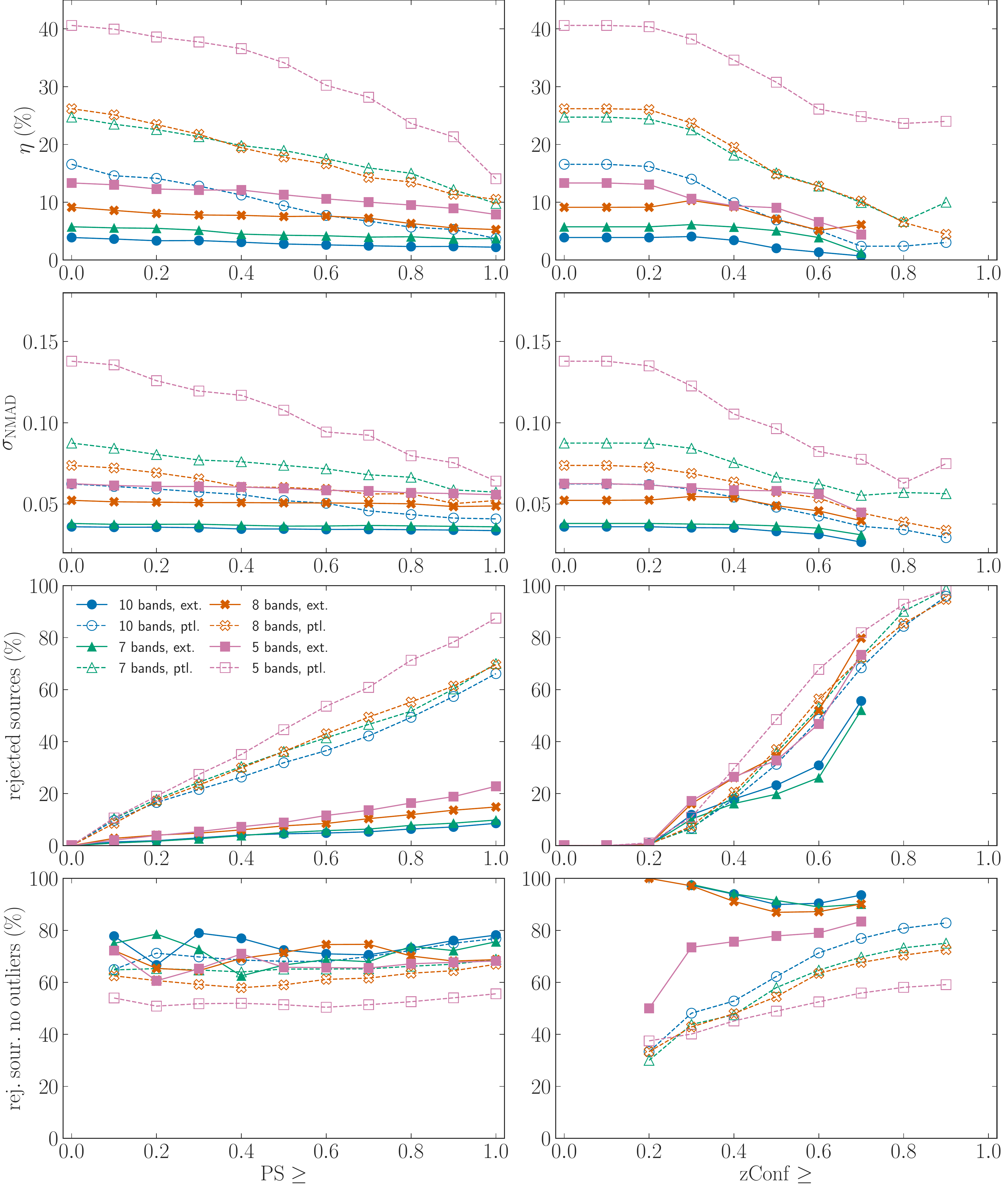}}
  \caption{Statistics for the derivation of photometric redshifts, 
  using the training+testing Pan-STARRS samples. All plots and symbols 
  as in Fig.~\ref{fig:zquality_arches}.}
\label{fig:zquality_pstarrs}
\end{figure*}

\subsection{Reliability and accuracy of the photometric redshifts}
\label{subsec:zquality}
\subsubsection{Statistical tests}
One of the main problems of photometric redshifts is estimating their 
accuracy and reliability. In the case of machine learning techniques, 
we can obtain a fine estimate of the method's performance, in the 
statistical sense, through tests using the corresponding training sample.

Each training sample presented in Sect.~\ref{subsec:training} was 
randomly split in two subsamples of equal size (training+testing). We 
used the training subsamples to estimate photometric redshifts with TPZ 
for the testing subsamples, and we compared these results with their 
corresponding spectroscopic redshifts. To this end we make use of the most 
widely used statistical indicators, which are the following:
\begin{align}
  x = \Delta(z_{norm}) & =  \frac{z_{spec}-z_{phot}}{1+z_{spec}}, \label{eq:zstats1}\\
  MAD(x) & =  Median(|x|), \\
  \sigma_\mathrm{NMAD}(x) & =  1.4826\times MAD(x), \\
  \eta & =  \frac{N_{outliers}}{N}\times 100, \label{eq:zstats4}
\end{align}
where $\sigma_\mathrm{NMAD}$ is the normalised median absolute deviation (MAD), 
and $\eta$ is the percentage of catastrophic outliers. A source is considered 
a catastrophic outlier if $|x|> 0.15$.

Table~\ref{tab:test1} (Cols.~1 and 4) shows the results of this test, comparing the 
performance of SDSS and Pan-STARRS training samples. Figure~\ref{fig:zphotzspec}
shows a comparison between spectroscopic and photometric redshifts for
SDSS (left) and Pan-STARRS (right) training samples.

Our test show that the Pan-STARRS training sample gets slightly better results 
for extended sources, but worse for point-like sources, particularly for 
photometric redshifts estimated using only optical data. It is worth noting 
that Pan-STARRS training samples for extended sources are a 20-30\%
smaller, but nevertheless their results are in fact better. The opposite 
happens with point-like sources: the SDSS training samples are about a
10\% smaller, but the performance is better. 

We note that this difference in the sizes of the extended and point-like 
training samples arise from the differences in the redshift distributions 
of SDSS and Pan-STARRS samples we mentioned above (Fig.~\ref{fig:zdist_training}). 
Extended sources in Pan-STARRS concentrate below redshift 1, while extended 
SDSS sources have a non negligible contribution of objects at redshift $\sim2$. 
The redshift distribution of the AGN population peaks at redshift two. We can 
therefore expect that a significant fraction of objects in the SDSS extended 
samples are AGN-dominated sources. Photometric redshifts for AGN-dominated 
galaxies are harder to estimate, due to variability effects and their roughly 
featureless, flat SED \citep[see e.g.][]{Salvato09}. This could be, partially at 
least, the reason behind the different performance between training samples.
In Pan-STARRS, most AGN-dominated sources are included in the point-like
sample, while host-dominated sources concentrates in the extended sample.
On the other hand, in the SDSS training samples, AGN- and host-dominated 
objects are more evenly distributed between extended and point-like samples.

We tested this possibility in two different ways. On one hand, we defined a 
new set of SDSS training samples (SDSS-b). We used the sources of the original
SDSS training samples, but only sources with $z<1$ were included in the 
extended training samples. The remaining sources were assigned to the 
corresponding point-like training samples. On the other hand, we again defined 
a new set of Pan-STARRS training samples (Pan-STARRS-b), but in this case 
using the SDSS classification to divide the original sample between extended 
and point-like sources. Then, we repeated again the same test we describe 
above for these new sets of training samples and estimated the corresponding 
statistics. 

Table~\ref{tab:test1} (Col.~2 and 5) shows the results for these tests. We
can see that removing high redshift sources from the SDSS extended samples
does in fact improve its performance, close to the levels of the Pan-STARRS
samples, or even better in some cases. The same effect is seen in the 
Pan-STARRS-b samples: the addition of high-$z$ objects to the extended 
samples indeed worsen the performance, although they are still better than 
the SDSS extended samples, except for the five filters sample.

However, this effect does not explain the poorer performance of the Pan-STARRS 
point-like samples. Both SDSS-b and Pan-STARRS-b point-like samples show
a slightly worse performance with respect to the original samples, but 
SDSS-b still obtained significantly better results.

Another significant difference between SDSS and Pan-STARRS training samples 
is the set of optical filters. To estimate how the presence of the u and y 
filters could affect the derivation of photometric redshifts, we defined two 
new sets of training samples (SDSS-c and Pan-STARRS-c), but using only the 
common filters between the SDSS and Pan-STARRS surveys (g, r, i, z). Then 
again, we repeated the same test by randomly splitting the samples in 
equal-sized training+testing samples. The results are presented in 
Table~\ref{tab:test1} (Cols.~3 and 6).

The effect of removing one optical filter in sources with NIR and/or MIR data
is not very significant, obtaining similar results in accuracy and fraction
of catastrophic outliers. For sources with only optical data the effect is more
profound, particularly for the SDSS training sample. The Pan-STARRS sample still
shows a better performance for extended sources, while the SDSS sample is better
for point-like sources, although the differences in performance between both
training samples are greatly reduced.

In principle, the relevance of each filter for the estimate of photometric 
redshifts can be highly dependant on the true redshift of the source. Among 
the ancillary products of TPZ there is an estimation of the relative 
importance of each attribute used to calculate the photometric redshifts. Figure~\ref{fig:uyimportance} shows the importance values we obtained for 
the u$-$g and i$-$y colours in our test using, respectively, the SDSS and 
Pan-STARRS five filters training samples. An importance near one means that 
the redshift information obtained through that particular colour is low 
compared with the rest of the colour set. This result shows that the y 
filter is relatively unimportant for the photometric redshifts in the 
Pan-STARRS sample, as the result of our previous test already suggested. It 
also shows how relevant is the u filter for sources with redshift above two.

Hence, this could explain, at least partially, the poorer performance of the 
Pan-STARRS training sample for point-like sources. As discussed above, this 
subsample contains a high fraction of high redshift sources, and therefore 
the lack of the u filter here can severely affects the calculation of 
photometric redshifts.

\subsubsection{PDF-based tests}
Since TPZ gives the full PDF for the photometric redshift in the 
predefined redshift space, we can obtain more information on the 
reliability of the derived redshift for each particular source 
\cite[see][]{tpz,Jones17}. An unimodal PDF, narrowly concentrated around 
its maximum is a sign of a reliable redshift estimate, while a 
multi-modal PDF with several local maxima of similar height is a clear 
sign that the redshift is badly determined.

We calculated several PDF-derived parameters and we tested how a 
selection of sources based on these parameters affects the statistics 
in our photo-z derivation. These parameters are defined as follows:

\begin{itemize}
\item zConf: The integral of the PDF in the interval 
      $\pm(1+z_\mathrm{phot})\times rms$, centred in $z_\mathrm{phot}$.
      $z_\mathrm{phot}$ is the mode of the PDF (the absolute maximum and the 
      value chosen as the photometric redshift of the source) and 
      $rms$ is the intrinsic dispersion of the method, which depends on 
      the employed training sample. For our sample we have used 
      $rms=0.06$ and it correspond to the root mean square of the out 
      of bag results given by TPZ (see \citealt{tpz}).
      A high value of zConf means that the PDF is highly concentrated 
      around $z_\mathrm{phot}$.
       
\item Npeaks: Number of local maxima (peaks) in the PDF.

\item PS (Peak strength): 1-P2/P1, where P1 is the probability density of 
      the highest local maximum in the PDF, and P2 is the second maximum peak.
      If the PDF is unimodal (P2=0) or P2$\ll$P1, $\mathrm{PS}\simeq 1$.
\end{itemize}

\begin{figure*}[t]
  \centering
  \resizebox{\hsize}{!}{\includegraphics{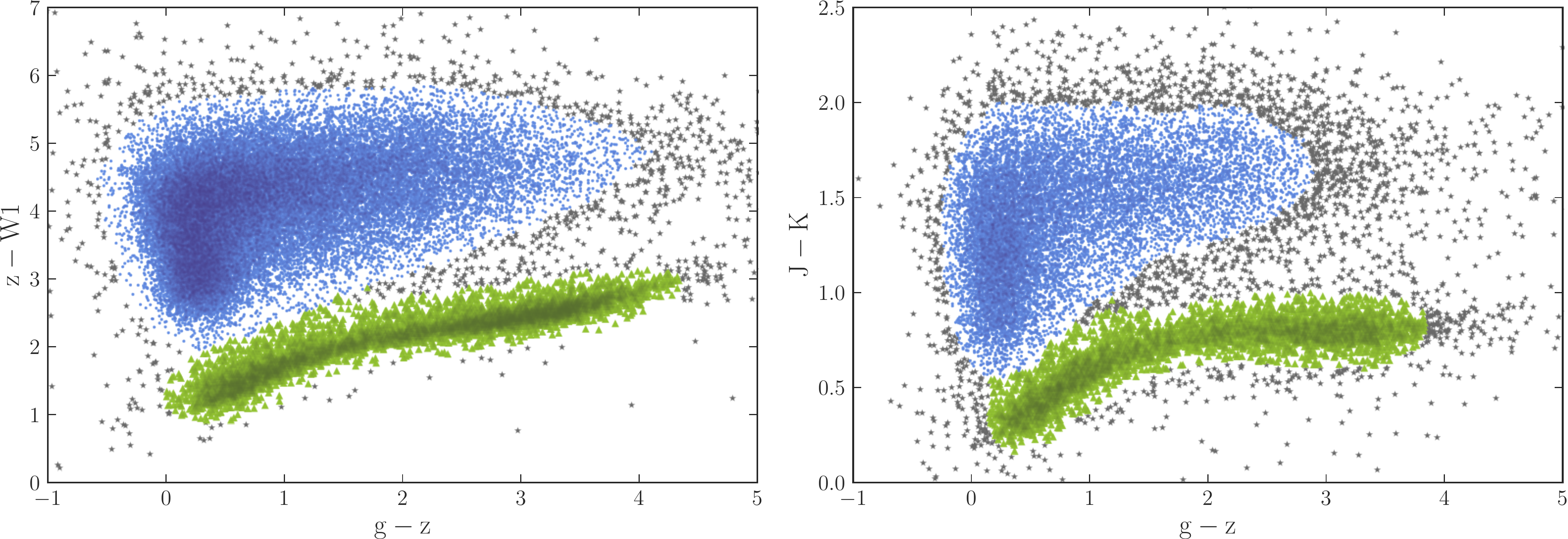}}
  \caption{Colour-colour plots of point-like XPS sources 
           showing our separation between stars and no-stars using 
           the hdbscan clustering algorithm (see Sect.~\ref{subsec:stars}). 
           \textbf{Left:} g$-$z versus z$-$W1. 
           \textbf{Right:} g$-$z versus J$-$K. 
           Green triangles are sources classified as stars; blue circles are 
           sources classified as no stars; grey asterisks show unclassified 
           objects. Darker shades mean a higher density of sources.}
\label{fig:starsplot1}
\end{figure*}

Figures~\ref{fig:zquality_arches} and \ref{fig:zquality_pstarrs} present 
the results of our statistical tests using the training+testing samples 
we described above, for the SDSS and Pan-STARRS samples, respectively.
For each sample, we selected sources with PS (left column) or zConf 
(right column) greater or equal than a given value, and we calculated 
the percentage of outliers, the normalised median absolute deviation, 
the percentage of rejected sources with respect to the total number of 
sources in the full sample (i.e. without any filtering), and the percentage 
of rejected sources that are not outliers with respect to the number of 
rejected sources. Npeaks is one if PS is one, therefore an Npeaks-based 
selection (Npeaks=1) is equivalent to selecting sources with PS=1.

There is an obvious trade-off between selecting high confidence photometric
redshifts and the final number of sources in the sample. Finding a compromise
between these two factors is therefore necessary. 

In the case of SDSS sources, our results show that a percentage of 
outliers below 10\% can be obtained for all but one samples, selecting 
sources with $\mathrm{PS}\geq 0.6$ or $\mathrm{zConf}\geq 0.5$. For the
five filters point-like sample both selections obtain a percentage of 
outliers about 20\%. The values of $\sigma_\mathrm{NMAD}$ are in both 
cases between 0.03 and 0.06, with slightly lower values for the zConf
selection. The zConf selection also gives lower percentages of outliers,
but the fraction of rejected sources is significantly higher for extended 
sources. With $\mathrm{PS}\geq 0.6$, about 10-20\% of 
extended sources are rejected, while imposing $\mathrm{zConf}\geq 0.5$
rejects about 30-50\%. In both cases the percentage of rejected point-like
sources is about 30-40\%.

For Pan-STARRS sources we found that a selection with $\mathrm{PS}\geq 0.7$ 
or $\mathrm{zConf}\geq 0.5$ obtains a percentage of outliers below 10\%
for extended sources and below 15\% for point-like sources. The values of 
$\sigma_\mathrm{NMAD}$ are between 0.03 and 0.07 in both cases. The exception
is the 5-filters point-like sample, with $\eta\sim 30\%$ and 
$\sigma_\mathrm{NMAD}\sim 0.09$. Although the percentage of outliers for 
this sample is still high, applying a quality filter significantly reduces 
this percentage. Again, the zConf selection gives a slightly lower percentage
of outliers, but with a significant higher loss of extended sources. With
$\mathrm{PS}\geq 0.7$ the percentage of rejected sources is less than 20\%
for extended sources and about 40-60\% for point-like sources. With 
$\mathrm{zConf}\geq 0.5$  the fraction of rejected point-like sources is 
similar, but it reaches 30-50\% for extended sources.

Note also that by imposing a quality cut we got a significant lose of
sources that are not outliers (see the last row panels of 
Figs.~\ref{fig:zquality_arches} and \ref{fig:zquality_pstarrs}). For the 
case of a PS selection, the fraction of rejected not-outliers is between
40 and 80\%, depending on the particular subsample, but it remains roughly
constant in spite of the selected PS value. On the other hand, the fraction 
for a zConf-based selection is much more dependant on the selected zConf 
value, varying between $\sim 20$ and 90\%.

Both PS and zConf values are included in our final catalogue, but we did not
apply any quality selection based on these parameters. It is up to the final 
users to find a compromise value that is best suited for their particular 
scientific goals.

\subsection{Identification of stars}
\label{subsec:stars}
At the average X-ray flux levels of the 3XMM catalogue and high
galactic latitudes, the expected percentage of X-ray emitting stars 
is small, below 10\%, but not negligible \citep{xms}. TPZ can be 
used also to classify sources, and it has in fact been used before 
to separate optical stars from quasars with an extremely high 
efficiency by using SDSS and WISE photometry \citep{Carrasco15}. 

As mentioned in Sect.~\ref{sec:photoz}, TPZ includes the option of
using classification trees while generating the random forest. We
can therefore use TPZ with a training sample where the sources were
classified in different categories. The number of possible categories
is in principle arbitrary, but for our case of star identification
we can simply use two values: one if the source is a star and zero 
if the source is not a star. The properties of this training sample 
(in our case, photometric colours) can be used to build classification
trees and a random forest in a way similar to that presented in 
Sect.~\ref{sec:photoz}. The resulting random forest can then be 
applied to different samples to separate sources between stars and
no stars.

Our first approach for identifying stars in XMMPZCAT was building a 
training sample of X-ray emitting stars with SDSS data. To this end, 
we retrieved the ROSAT all-sky survey and SDSS sample of X-ray 
emitting stars \citep[RASSDSSTAR;][]{rassdsstar}. However the size 
of the sample we were able to build using RASSDSSTAR was quite low, 
$\sim$ 700 sources, and this number was even lower for sources with 
Pan-STARRS data ($\lesssim 500$).

Hence we used a different approach. Stars are easily separated from 
galaxies and QSO using a combination of optical and IR colours (see
e.g. \citealt{Wu12}). The dust content of most stars is low, so it is 
expected that most stars show lower IR colours (e.g. z$-$W1 or J$-$K) 
than galaxies or QSO. We can use the objects included in our cross-matched
catalogues (ARCHES or XPS) having IR photometry for selecting X-ray 
emitting stars through colour-colour plots. 

In the ARCHES (XPS) sample we have 16\,321 (50\,614) point-like 
sources with the needed optical and IR colours. Figure~\ref{fig:starsplot1} 
shows two colour-colour plots for XPS objects. On the left we plotted 
g$-$z versus z$-$W1 and on the right g$-$z versus J$-$K. As expected, stars 
are located in the lower region of the plots. For a systematic separation 
between stars and galaxies/QSO we applied HDBSCAN* \citep{Campello13}, a hierarchical, 
density-based clustering algorithm, as implemented in the python 
package \texttt{hdbscan}\footnote{\url{https://github.com/scikit-learn-contrib/hdbscan}} 
\citep{McInnes17}.

\begin{figure}[t]
  \centering
  \resizebox{\hsize}{!}{\includegraphics{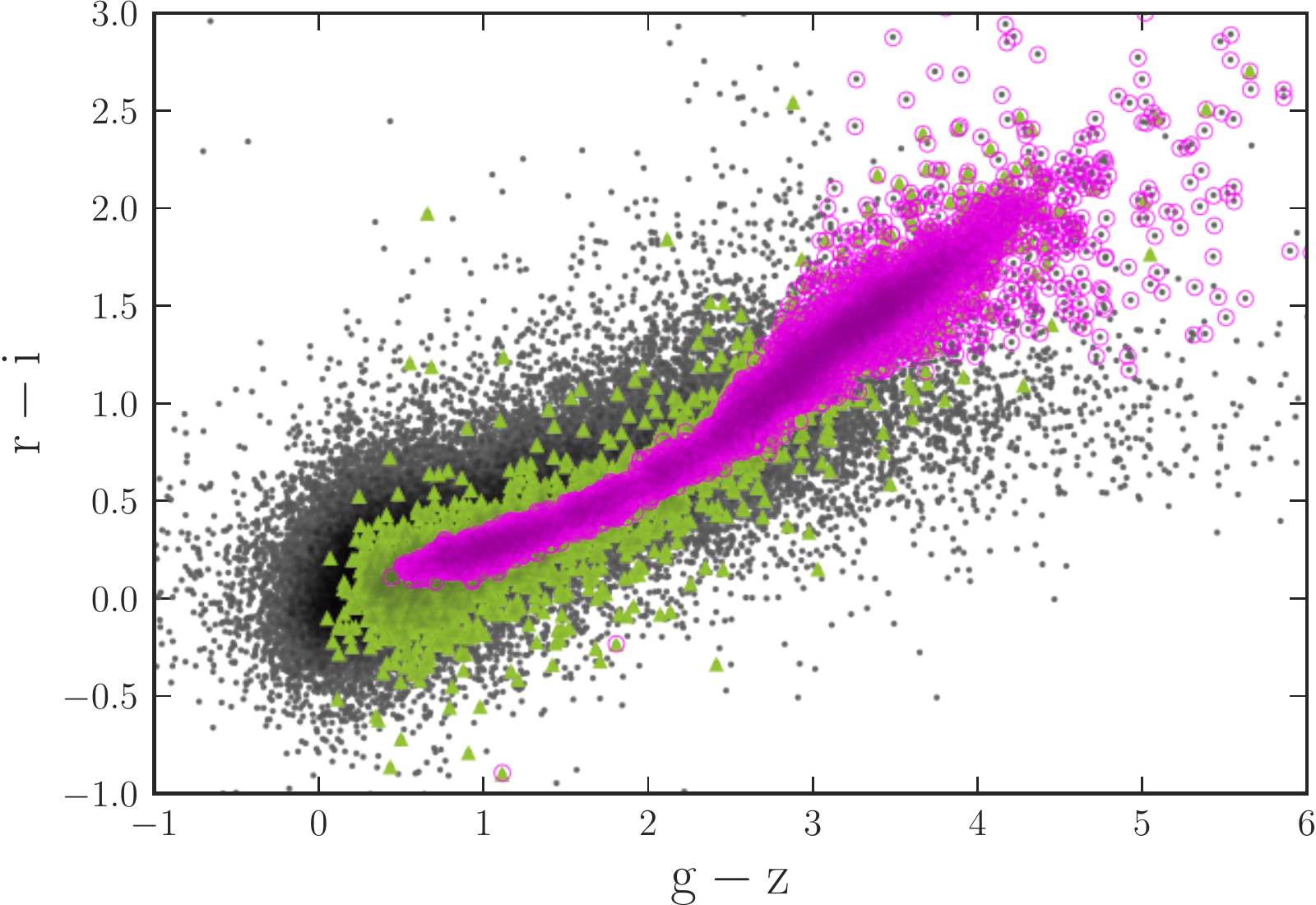}}
  \caption{r$-$i versus g$-$z colour-colour plot of point-like 
           XPS sources. Magenta open circles are sources identified as 
           stars using TPZ. Green triangles are sources classified as stars using 
           colour criteria (Fig.~\ref{fig:starsplot1}). Grey solid circles are the 
           remaining point-like objects in the Pan-STARRS sample. Darker shades 
           mean a higher density of sources.}
\label{fig:starsplot2}
\end{figure}

We run \texttt{hdbscan} for each colour-colour plot presented in 
Fig.~\ref{fig:starsplot1}, dividing the sample between stars (green 
triangles) and galaxies/QSO (blue circles). Grey asterisks are sources 
that remained unclassified using the hdbscan algorithm. We imposed 
a minimum cluster size of 100 objects, and we kept the rest of 
parameters of the algorithm at the default values. 

In order to build our training sample for TPZ, we selected as stars 
objects identified as stars using NIR or MIR colours, and as no-stars 
objects identified as galaxies/QSO in NIR and MIR colours. Applying 
these criteria we build a training sample of 2816 stars and 10\,992 
no-stars for the ARCHES sample, and of 5555 stars and 36\,004 no-stars 
for the XPS sample.

Using these training samples in TPZ we can identify stars for those
sources with only optical colours available. For the ARCHES (XPS) 
sample TPZ identified 4680 (5636) objects as stars, 2008 (1599) of 
them not identified through the IR-colour selection. In the final catalogue 
of 100\,178 X-ray sources there are 10\,830 stars ($\sim 11\%$), after 
combining those stars identified through IR colours and those identified 
using TPZ. Figure~\ref{fig:starsplot2} shows an optical colour-colour 
diagram for point-like Pan-STARRS sources. The plot clearly shows that, 
using our training sample, TPZ is able to identify the typical star tail 
in this kind of diagrams (magenta open circles).

We used this information to flag sources as stars in XMMPZCAT (see 
Appendix). We included those sources that were identified as stars 
using their IR colours and the hdbscan algorithm, and those identified
through their optical colours using TPZ.

\begin{figure*}
  \begin{minipage}{0.5\textwidth}
    \centering
    \resizebox{\hsize}{!}{\includegraphics{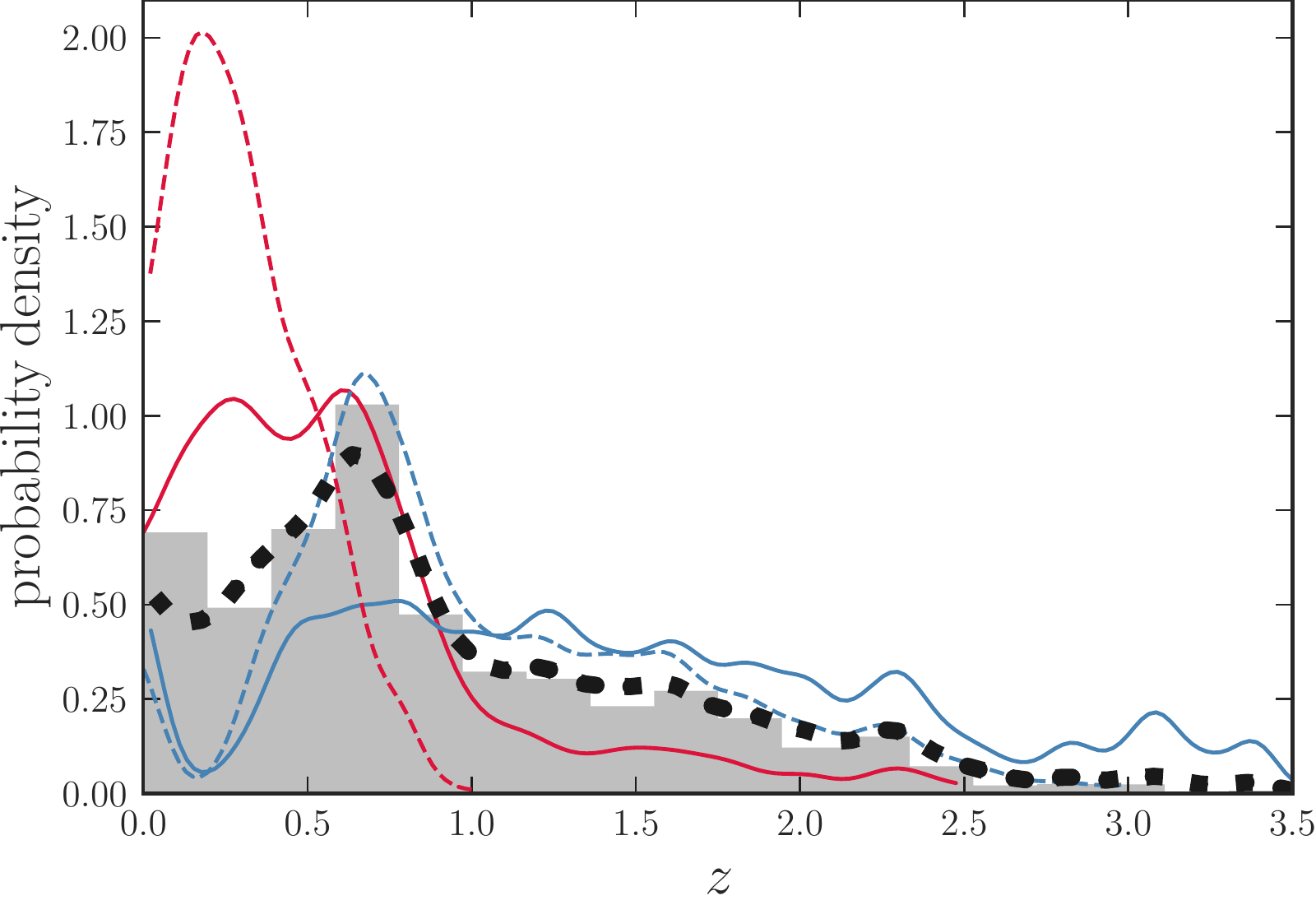}}
  \end{minipage}
  \begin {minipage}{0.5\textwidth}
    \centering
    \resizebox{\hsize}{!}{\includegraphics{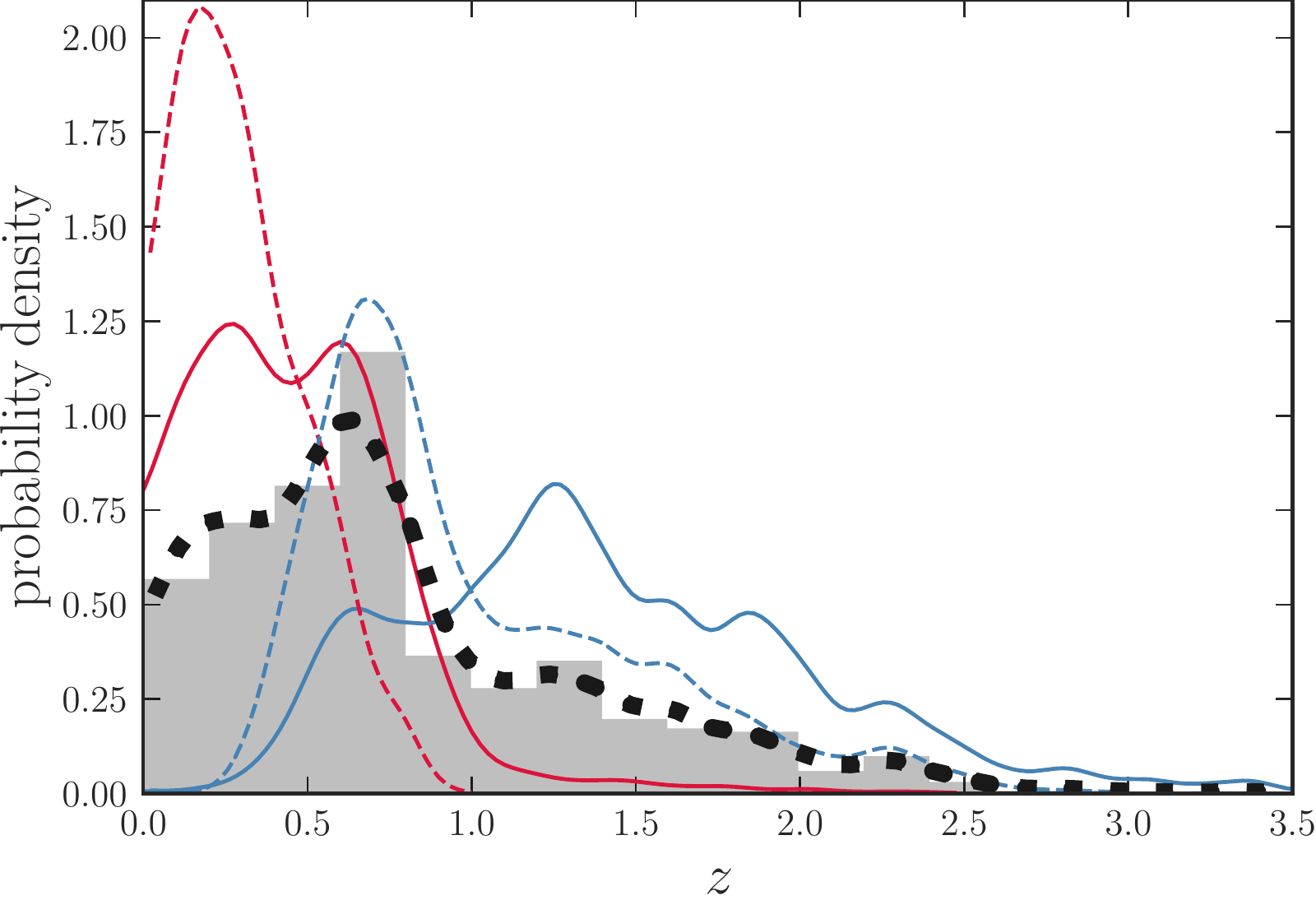}}
  \end{minipage}
   
  \caption{Normalised redshift distributions of XMMPZCAT sources. 
           \textbf{Left:} Full catalogue, with no filtering. 
           \textbf{Right:} Sources with peak strength greater or equal than 0.7 
           (see Sect.~\ref{subsec:zquality}) and not classified as stars (see 
           Sect.~\ref{subsec:stars}). The filled histogram shows the normalised 
           distribution using the most probable redshift (the mode) for each 
           source. Lines show the probability density of the sample estimated 
           by adding all the corresponding PDFs. Red lines correspond to extended 
           sources, blue lines to point-like sources. Solid lines represent 
           ARCHES samples, dashed lines the Pan-STARRS samples. Dotted, black line
           represents the whole sample.}
\label{fig:zdist}   
\end{figure*}

\begin{figure*}
  \centering
  \resizebox{\hsize}{!}{\includegraphics{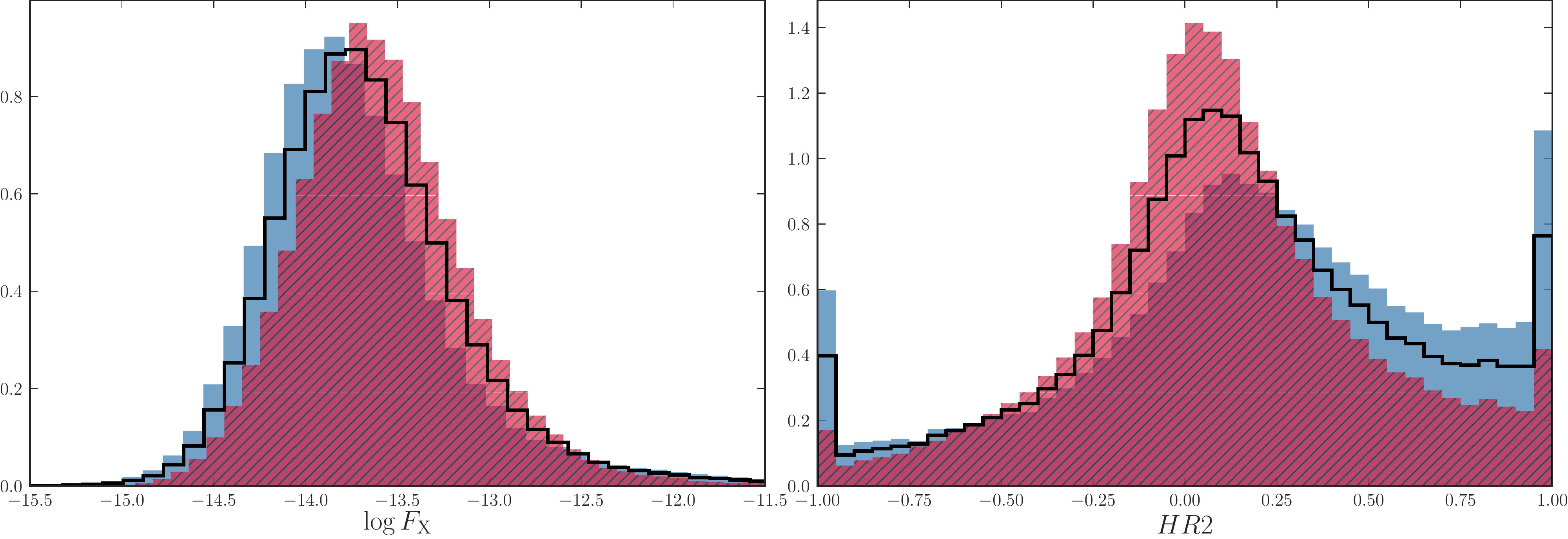}}   
  \caption{X-ray properties of XMMPZCAT sources (red, hatched histogram), compared 
           with the total population of 3XMM sources in the XMMPZCAT footprint
           (black, solid line) and the population of 3XMM sources in the XMMPZCAT
           footprint but with no estimated photo-z (blue, solid histogram).
           \textbf{Left:} Normalised 0.2-12~keV flux distributions. 
           \textbf{Right:} Normalised hardness ratio (HR2) distributions.}
\label{fig:fxdist}   
\end{figure*}

\subsection{Properties of XMMPZCAT sources}
\label{subsec:catprop}

In this section we present some of the global properties of XMMPZCAT 
sources, such as redshift, X-ray flux and hardness ratio distributions. 

Figure~\ref{fig:zdist} shows the redshift distribution before (left) 
and after (right) filtering by PDF peak strength (PS$\geq 0.7$), for 
the full catalogue (black dotted lines) and for each subsample: 
extended or point-like (red or blue lines), and ARCHES or XPS (solid or 
dashed lines). We estimated the distributions by adding the TPZ-estimated 
PDF of each source in the catalogue. For comparison, we also plotted the 
total redshift distribution estimated using the final photo-z values (the 
PDF's mode) included in the catalogue (grey histogram).

Our results show that most XMMPZCAT sources are concentrated below 
redshift $\sim 1$, with a strong peak at $z \sim 0.7$ and a long tail 
up to redshift $\sim 2.5$. XPS extended sources show a strong peak 
at $z \sim 0.2$, with no significant contribution for redshifts greater 
than 1. Most XPS point-like objects accumulate between redshift 
0.3 and 1, with a strong peak at $z \sim 0.7$, a small hump at $z \sim 1.3$ 
and a long tail up to redshift 2.5. 

On the other hand, as discussed in Sect.~\ref{subsec:training}, SDSS can 
detect extended sources at higher redshifts, and hence ARCHES samples show 
a different distribution between extended and point-like sources. The 
redshift distribution of the ARCHES extended sample show two peaks at 
$z \sim 0.2$ and $\sim 0.7$. ARCHES point-like sources are more abundant 
at higher redshifts than XPS point-like objects, with a peak around 
$\sim 1.3$ and broadly distributed between redshift $\sim 0.5$ and $\sim 2.5$.

Figure~\ref{fig:fxdist} shows the 0.2-12~keV flux (left) and hardness 
ratio (right) distributions for the XMMPZCAT (red histogram). For 
comparison we have also included two samples of 3XMM sources in the 
XMMPZCAT footprint: the black solid line shows the distribution for all 
3XMM sources in the footprint and the blue histogram shows only those with
no estimated photo-z. We used the hardness ratio between 0.5-1 and 1-2~keV
as given in the 3XMM catalogue \citep[HR2, see][]{3xmm}.\footnote{The 
hardness ratio is defined as $HR(A,B)=(B-A)/(A+B)$, where $A$ and $B$ 
are the X-ray counts at two different energy bands. The energy of the B
band is higher than the energy of the A band.} If a source has a hardness
ratio close to one, it means that its X-ray emission is larger at high 
energies than at low energies. Since low energies are more sensitive to
X-ray absorption, high hardness ratios are usually understood as a sign of
obscuration \citep[see e.g.][]{Severgnini12}.

The distribution of XMMPZCAT objects is skewed towards higher X-ray fluxes, 
compared with the total 3XMM population. The median X-ray flux for XMMPZCAT
sources is $2.4 \times 10^{-14}~\mathrm{erg\,s^{-1}\,cm^{-2}}$ and 
$1.6 \times 10^{-14}~\mathrm{erg\,s^{-1}\,cm^{-2}}$ for 3XMM sources with 
no photo-z. A Kolmogorov-Smirnov test shows that these distributions are
different with $\gg 5\sigma$ confidence. 

XMMPZCAT sources seem to present a softer X-ray spectral shape. The hardness 
ratio distribution is skewed towards lower HR2 values, with a median of 0.08, 
while 3XMM sources with no photo-z show a median of 0.2. There is a clear 
under-abundance of XMMPZCAT sources with high hardness ratios compared with the
total 3XMM population.

Higher X-ray fluxes and softer spectral shapes suggest that, on average, 
XMMPZCAT sources could show a lower X-ray obscuration than the global 
3XMM population. Although a more detailed analysis of the X-ray properties 
of these objects would be interesting, this is probably just a selection
effect, since we are looking for X-ray sources with counterparts
with good photometry.

\section{Conclusions}
\label{sec:conclusions}
We presented XMMPZCAT, a catalogue of photometric redshifts for XMM-\textit{Newton}
serendipitous X-ray sources. Using TPZ, a machine learning algorithm based
on decision trees and random forest, and the training sample presented in
\citet{Mountrichas17}, we estimated photometric redshifts for 100\,178 objects
in the 3XMM-DR6 catalogue outside the Galactic plane. The base of our photo-z
estimations is optical photometry from the SDSS and Pan-STARRS catalogues,
with additional photometry in the NIR (2MASS, UKIDSS, VISTA-VHS) and MIR (AllWISE) 
bands if available. XMMPZCAT contains photo-z for about 50\% of the X-ray 
sources within the XMM-\textit{Newton} observations selected to build the catalogue. 

We tested the reliability and accuracy of these photo-z using statistical 
tests based on the training samples. We found that the accuracy of our results is
highly dependent on the number of available filters and the extension (i.e. if 
the optical counterpart is extended or point-like) of the source. The fraction 
of outliers range between 4\% (9\%) for extended Pan-STARRS (ARCHES) sources 
with 10 available filters, up to 40\% (30\%) for point-like Pan-STARRS (ARCHES) 
objects with only optical photometry.

We defined three parameters (PS, zConf, Npeaks) that describe the overall shape 
of the photo-z PDF for each source. These parameters show how reliable is the 
estimated photo-z. We found that a selection of sources such as PS>0.6 
significantly reduced the fraction of outliers, particularly for point-like 
objects. However, no filtering based on this kind of quality criteria was 
applied in the final version of XMMPZCAT. We included our estimates of this
parameters in the catalogue, and is up to the final user to apply the quality
selection best suited for the pursued scientific goals.

We also identified the content of stars in the catalogue. We build a training 
sample of IR colour-selected stars and used TPZ to identify stars among objects
with only optical colours. We found that about 10\% of the sources are stars, a
result consistent with other high Galactic latitude X-ray surveys .

Our results show that most XMMPZCAT sources are concentrated below 
redshift $\sim 1$, with a strong peak at $z \sim 0.7$ and a long tail 
up to redshift $\sim 2.5$. They seem to be, on average, slightly brighter
in X-rays and with a softer X-ray spectral shape than the total population
of 3XMM sources (most probably a selection effect because of searching for
counterparts with good photometry).

XMMPZCAT can nevertheless be improved in the future. The size of the 
catalogue can grow with future versions of 3XMM and the release of new,
deeper large area optical surveys. We will also be able to increase the 
accuracy and reliability of the photo-z: for example by assembling larger 
training samples, by improving our cross-matching techniques including the 
photometric data \citep{Salvato18} or through variability corrections 
\citep{Wolf04} using the Pan-STARRS multi-epoch information. The latest 
version of the catalogue is always available in the web site of this project.\footnote{\url{http://xraygroup.astro.noa.gr/Webpage-prodex/index.html}}

XMMPZCAT is a key step to the full exploitation of the scientific potential of
the 3XMM catalogue. Thanks to the distance information we have derived, we can
improve other 3XMM added-value products. For example, using photometric redshifts
along XMMFITCAT we can derive important properties of X-ray sources like X-ray 
luminosity, temperature of the hot emitting plasmas, rest-frame obscuring column 
densities and rest-frame energies of emission and absorption features.

\begin{acknowledgements}
This work is part of the Enhanced XMM-Newton Spectral-fit Database project, 
funded by the European Space Agency (ESA) under the PRODEX program.
A.C.R. acknowledge financial support through grant AYA2015-64346-C2-1-P (MINECO/FEDER).
G.M. acknowledges financial support from the AHEAD project, which is funded by 
the European Union as Research and Innovation Action under Grant No: 654215.
\\
This research has made use of data obtained from the 3XMM XMM-\textit{Newton} 
serendipitous source catalogue compiled by the ten institutes of the 
XMM-\textit{Newton} Survey Science Centre selected by ESA.
\\
This work is based on observations made with XMM-\textit{Newton}, an ESA science 
mission with instruments and contributions directly funded by ESA Member States 
and NASA. 
\\
Funding for the Sloan Digital Sky Survey IV has been provided by the Alfred P. Sloan Foundation, the U.S. Department of Energy Office of Science, and the Participating Institutions. SDSS-IV acknowledges
support and resources from the Center for High-Performance Computing at
the University of Utah. The SDSS web site is \url{www.sdss.org}.
\\
SDSS-IV is managed by the Astrophysical Research Consortium for the 
Participating Institutions of the SDSS Collaboration including the 
Brazilian Participation Group, the Carnegie Institution for Science, 
Carnegie Mellon University, the Chilean Participation Group, the French Participation Group, Harvard-Smithsonian Center for Astrophysics, 
Instituto de Astrof\'isica de Canarias, The Johns Hopkins University, 
Kavli Institute for the Physics and Mathematics of the Universe (IPMU) / 
University of Tokyo, Lawrence Berkeley National Laboratory, 
Leibniz Institut f\"ur Astrophysik Potsdam (AIP),  
Max-Planck-Institut f\"ur Astronomie (MPIA Heidelberg), 
Max-Planck-Institut f\"ur Astrophysik (MPA Garching), 
Max-Planck-Institut f\"ur Extraterrestrische Physik (MPE), 
National Astronomical Observatories of China, New Mexico State University, 
New York University, University of Notre Dame, 
Observat\'ario Nacional / MCTI, The Ohio State University, 
Pennsylvania State University, Shanghai Astronomical Observatory, 
United Kingdom Participation Group,
Universidad Nacional Aut\'onoma de M\'exico, University of Arizona, 
University of Colorado Boulder, University of Oxford, University of Portsmouth, 
University of Utah, University of Virginia, University of Washington, University of Wisconsin, 
Vanderbilt University, and Yale University.
\\
The Pan-STARRS1 Surveys (PS1) and the PS1 public science archive have been made possible through contributions by the Institute for Astronomy, the University of Hawaii, the Pan-STARRS Project Office, the Max-Planck Society and its participating institutes, the Max Planck Institute for Astronomy, Heidelberg and the Max Planck Institute for Extraterrestrial Physics, Garching, The Johns Hopkins University, Durham University, the University of Edinburgh, the Queen's University Belfast, the Harvard-Smithsonian Center for Astrophysics, the Las Cumbres Observatory Global Telescope Network Incorporated, the National Central University of Taiwan, the Space Telescope Science Institute, the National Aeronautics and Space Administration under Grant No. NNX08AR22G issued through the Planetary Science Division of the NASA Science Mission Directorate, the National Science Foundation Grant No. AST-1238877, the University of Maryland, Eotvos Lorand University (ELTE), the Los Alamos National Laboratory, and the Gordon and Betty Moore Foundation.
\\
This publication makes use of data products from the Wide-field Infrared Survey 
Explorer, which is a joint project of the University of California, Los Angeles, 
and the Jet Propulsion Laboratory/California Institute of Technology, funded by 
the National Aeronautics and Space Administration. 
\\
The UKIDSS project is defined in \citealt{ukidss}. UKIDSS uses the UKIRT 
Wide Field Camera \citep[WFCAM;][]{Casali07} and a photometric system described 
in \citealt{Hewett06}. The pipeline processing and science archive are described 
in \citealt{Irwin04} and \citealt{Hambly08}. We have used data from the tenth 
data release.
\\
The VISTA Data Flow System pipeline processing and science archive are described 
in \citealt{Irwin04}, \citealt{Hambly08} and \citealt{Cross12}. Based on 
observations obtained as part of the VISTA Hemisphere Survey, ESO Program, 
179.A-2010 (PI: McMahon). We have used data from the third data release.
\\
This publication makes use of data products from the Two Micron All Sky Survey, 
which is a joint project of the University of Massachusetts and the Infrared 
Processing and Analysis Center/California Institute of Technology, funded by the 
National Aeronautics and Space Administration and the National Science 
Foundation.
\\
This research made use of Astropy, a community-developed core Python package 
for Astronomy \citep{astropy2}.
\end{acknowledgements}

%-------------------------------------------------------------------

\bibliographystyle{aa}
\bibliography{astrobib}

\begin{appendix}
\section*{Appendix: Description of XMMPZCAT}
\label{catalogue}
\balance
XMMPZCAT consists of a FITS table with one row for each unique X-ray source, and 
17 columns containing the estimated redshift plus additional information about 
the X-ray source, the optical counterpart and several parameters that can help 
assessing the reliability of the derived photometric redshift. Not available values 
are represented by a ``null'' value. The columns of this table correspond to:

\begin{itemize}
\item{{\tt XMM\_SRCID}: Source identification number as in 3XMM-DR6.}
\item{{\tt XMM\_RA, XMM\_DEC}: X-ray source coordinates as in 3XMM-DR6.}
\item{{\tt XMMFITCAT}: True if the source is in the XMMFITCAT.}
\item{{\tt XMATCH}: Origin of the optical counterpart (SDSS-ARCHES or Pan-STARRS).}
\item{{\tt proba\_XMATCH}: Probability that all counterparts are associated 
with the same real source, as estimated by \texttt{xmatch}.}
\item{{\tt opt\_SRCID}: Source identification number in SDSS-DR13 or Pan-STARRS-DR1.}
\item{{\tt Nfilters}: Number of photometric data used.}
\item{{\tt extended}: True if the sources is extended in SDSS-DR13, false if 
it is point-like.}
\item{{\tt ph\_flag}: Quality of the photometric data (see note below).} 
\item{{\tt inTCS} (in Training Colour Space): True if all colours used to 
calculate the photometric redshift are inside the colour space well covered by 
the corresponding training sample.}
\item{{\tt STARS}: True if the source was identified as a star. It includes both
sources identified using IR colours and hdbscan, and optical colours and TPZ.}
\item{{\tt SPEC\_Z}: Spectroscopic redshift in SDSS-DR13 (null if not available).}
\item{{\tt PHOT\_Z}: Derived photometric redshift.}
\item{{\tt PHOT\_ZERR}: One-sigma error of the derived photometric redshift.}
\item{{\tt PHOT\_ZCONF}: Confidence of the photometric redshift. It gives an 
idea about how narrowly concentrated is the redshift probability distribution 
(PDF) around PHOT\_Z.}
\item{{\tt Npeaks}: Number of local maxima (peaks) in the PDF.}
\item{{\tt PS} (Peak Strength): $1-P2/P1$, where $P1$ is the probability density 
of the highest local maximum in the PDF, and $P2$ is the second maximum peak.}
\item{{\tt PHOT\_Z2}:  redshift position of $P2$.}
\end{itemize}

\textbf{Note on {\tt ph\_flag}.}  The values of this column have a three character 
format, XYZ, where X is the flag for optical data, Y for WISE data and Z for 
NIR data (2MASS, UKIDSS or VISTA). The quality of the data was obtained from the
corresponding flags of the respective catalogues. The possible values for X/Y/Z are:
\begin{itemize}
\item{A: all magnitudes in this range are flagged as good.}
\item{B: some magnitudes in this range have bad photometry.}
\item{F: all magnitudes are bad.}
\item{0: no photometric data in this range.}
\end{itemize}

In addition, a supplementary FITS table is provided, containing the 
estimated probability density distribution for the photometric redshift 
of each source. Both tables can be downloaded from the web site of the 
project.\footnote{\url{http://xraygroup.astro.noa.gr/Webpage-prodex/xmmpzcat_access.html}} 

\end{appendix}
\end{document}